\documentclass[twocolumn]{aastex631}

\usepackage{natbib}
\usepackage{amsmath}
\usepackage{subfigure}
\setcitestyle{sort,comma}
\usepackage{csvsimple}

\usepackage[version=4]{mhchem} 
\usepackage{seqsplit}

\newcommand{\mum}{\ensuremath{\mathrm{\,\mu m}}}

\newcommand{\Zatm}{\ensuremath{Z_{\mathrm{atm}}}}
\newcommand{\Zenv}{\ensuremath{Z_{\mathrm{env}}}}

\newcommand{\logX}[1]{\ensuremath{\log(\mathrm{X_{\ce{#1}}})}}
\newcommand{\logXratio}[2]{\ensuremath{\log(\mathrm{X_{\ce{#1}} / X_{\ce{#2}} })}}

\newcommand{\re}{\,R_\oplus}
\newcommand{\me}{\,M_\oplus}

\providecommand{\apjl}{ApJL}
\providecommand{\apjs}{ApJS}
\providecommand{\aap}{A\&Aj}

\graphicspath{{./}{figures/}}

\begin{document}

\title{\large{\textit{JWST} Reveals CH$_4$, CO$_2$, and H$_2$O in a Metal-rich Miscible Atmosphere\\ on a Two-Earth-Radius Exoplanet}}

\newcommand{\umontreal}{Department of Physics and Trottier Institute for Research on Exoplanets, Universit\'{e} de Montr\'{e}al, Montreal, QC, Canada \href{mailto:bjorn.benneke@umontreal.ca}{bjorn.benneke@umontreal.ca}}

\author[0000-0001-5578-1498]{Bj\"{o}rn Benneke} 
\affil{\umontreal}

\author[0000-0001-6809-3520]{Pierre-Alexis Roy} 
\affil{\umontreal}

\author[0000-0002-2195-735X]{Louis-Philippe Coulombe} 
\affil{\umontreal}

\author[0000-0002-3328-1203]{Michael Radica} 
\affil{\umontreal}

\author[0000-0002-2875-917X]{Caroline Piaulet}
\affil{\umontreal}

\author[0000-0003-0973-8426]{Eva-Maria Ahrer}
\affil{Max Planck Institute for Astronomy, K\"{o}nigstuhl 17, D-69117 Heidelberg, Germany}

\author[0000-0002-5887-1197]{Raymond Pierrehumbert}
\affil{University of Oxford, Department of Physics Oxford, OX1 3PW, UK}

\author[0000-0001-6878-4866]{Joshua Krissansen-Totton}
\affil{Department of Earth and Space Sciences/Astrobiology Program, University of Washington, Seattle, WA, USA}

\author[0000-0002-0298-8089]{Hilke E. Schlichting}
\affil{Department of Earth, Planetary, and Space Sciences, University of California, Los Angeles, Los Angeles, CA 90095, USA}

\author[0000-0003-2215-8485]{Renyu Hu}
\affil{Jet Propulsion Laboratory, California Institute of Technology, Pasadena, CA 91109, USA}
\affil{6 Division of Geological and Planetary Sciences, California Institute of Technology, Pasadena, CA 91125, USA}

\author[0000-0002-1551-2610]{Jeehyun Yang}
\affil{Jet Propulsion Laboratory, California Institute of Technology, Pasadena, CA 91109, USA}

\author[0000-0002-4997-0847]{Duncan Christie}
\affil{Max Planck Institute for Astronomy, K\"{o}nigstuhl 17, D-69117 Heidelberg, Germany}

\author[0000-0002-5113-8558]{Daniel Thorngren}
\affil{Department of Physics and Astronomy, Johns Hopkins University}

\author[0000-0002-1299-0801]{Edward D. Young}
\affil{Department of Earth, Planetary, and Space Sciences, University of California, Los Angeles, Los Angeles, CA 90095, USA}

\author[0000-0002-8573-805X]{Stefan Pelletier} 
\affil{Observatoire astronomique de l'Universit\'{e} de Gen\`{e}ve, 51 chemin Pegasi 1290 Versoix, Switzerland}
\affil{\umontreal}

\author[0000-0002-5375-4725]{Heather A. Knutson}
\affil{Division of Geological and Planetary Sciences, California Institute of Technology, Pasadena, CA 91125, USA}

\author[0000-0002-0747-8862]{Yamila Miguel}
\affil{Leiden Observatory, Leiden University, P.O. Box 9513, 2300 RA Leiden, The Netherlands}
\affil{SRON Netherlands Institute for Space Research, Niels Bohrweg 4, 2333 CA Leiden, The Netherlands}

\author[0000-0001-5442-1300]{Thomas M. Evans-Soma}
\affiliation{School of Information and Physical Sciences, University of Newcastle, Callaghan, NSW, Australia}
\affiliation{Max Planck Institute for Astronomy, K\"{o}nigstuhl 17, D-69117 Heidelberg, Germany}

\author[0000-0001-6110-4610]{Caroline Dorn}
\affil{Department of Physics and Institute for Particle Physics and Astrophysics, ETH Zurich, CH-8093 Zurich, Switzerland}

\author[0009-0003-2576-9422]{Anna Gagnebin}
\affil{Department of Astronomy \& Astrophysics, University of California, Santa Cruz, CA 95064, USA}

\author[0000-0002-9843-4354]{Jonathan J. Fortney}
\affil{Department of Astronomy \& Astrophysics, University of California, Santa Cruz, CA 95064, USA}

\author[0000-0002-9258-5311]{Thaddeus Komacek}
\affil{Department of Astronomy, University of Maryland, College Park, MD 20742, USA}

\author[0000-0003-4816-3469]{Ryan MacDonald}
\affil{Department of Astronomy, University of Michigan, Ann Arbor, MI, USA}

\author[0009-0002-2380-6683]{Eshan Raul}
\affil{Department of Astronomy, University of Michigan, Ann Arbor, MI, USA}

\author[0000-0001-5383-9393]{Ryan Cloutier}
\affil{Department of Physics \& Astronomy, McMaster University, 1280 Main St W, Hamilton, ON, L8S 4L8, Canada}

\author[0000-0002-9147-7925]{Lorena Acuna}
\affil{Max Planck Institute for Astronomy, K\"{o}nigstuhl 17, D-69117 Heidelberg, Germany}

\author[0000-0002-6780-4252]{David Lafrenière}
\affil{\umontreal}

\author[0000-0001-9291-5555]{Charles Cadieux}
\affil{\umontreal}

\author[0000-0001-5485-4675]{René Doyon}
\affil{\umontreal}

\author[0000-0003-0156-4564]{Luis Welbanks}
\affil{School of Earth \& Space Exploration, Arizona State University, Tempe, AZ 85257, USA}

\author[0000-0002-1199-9759]{Romain Allart}
\affil{\umontreal}

\begin{abstract}
Even though sub-Neptunes likely represent the most common outcome of planet formation, their natures remain poorly understood. In particular, planets near 1.5--2.5$\re$ often have bulk densities that can be explained equally well with widely different compositions and interior structures, resulting in grossly divergent implications for their formation and potential habitability. Here, we present the full 0.6--5.2$\,\mu \mathrm{m}$ \textit{JWST NIRISS/SOSS}+\textit{NIRSpec/G395H} transmission spectrum of the 2.2$\,\re$ planet TOI-270\,d ($4.78\,\me$, $T_\mathrm{eq}$=350--380\,K), delivering unprecedented sensitivity for atmospheric characterization in the sub-Neptune regime. We detect five vibrational bands of CH$_4$ at 1.15, 1.4, 1.7, 2.3, and 3.3~$\mu$m (9.4$\sigma$), the signature of CO$_2$ at 4.3~$\mu$m (4.8$\sigma$), water vapor (2.5$\sigma$), and potential signatures of SO$_2$ at 4.0$\,\mu \mathrm{m}$ and CS$_2$ at 4.6$\,\mu\mathrm{m}$. Intriguingly, we find an overall highly metal-rich atmosphere, with a mean molecular weight of $5.47_{-1.14}^{+1.25}$. We infer an atmospheric metal mass fraction of $58_{-12}^{+8}\%$ and a C/O of $0.47_{-0.19}^{+0.16}$, indicating that approximately half the mass of the outer envelope is in high-molecular-weight volatiles (H$_2$O, CH$_4$, CO, CO$_2$) rather than H$_2$/He. We introduce a sub-Neptune classification scheme and identify TOI-270\,d as a ``miscible-envelope sub-Neptune'' in which H$_2$/He is well-mixed with the high-molecular-weight volatiles in a miscible supercritical metal-rich envelope. For a fully miscible envelope, we conclude that TOI-270\,d's interior is $90_{-4}^{+3}$\,wt\,\% rock/iron, indicating that it formed as a rocky planet that accreted a few wt\,\% of H$_2$/He, with the overall envelope metal content explained by magma-ocean/envelope reactions without the need for significant ice accretion. TOI-270\,d may well be an archetype of the overall population of sub-Neptunes.
\end{abstract}

\section{Introduction}\label{sec:intro}
One of the most fundamental results in the study of exoplanets has been the discovery that small close-in planets are bifurcated into two distinct populations: the super-Earths around 1.3 R$_\oplus$, and the sub-Neptunes around 2.4 R$_\oplus$   \citep{fulton_california-keplersurvey._2017, fulton_california-kepler_2018, vaneylen_asteroseismic_2018, hardegree-ullman_kepler_2019}. Separated by the radius valley near $1.8\re$, these planets have no analogues in the solar system, and present an intriguing problem from a planet formation standpoint. 

The first wave of spectroscopic studies of sub-Neptunes with the Hubble Space Telescope (HST) found growing evidence that the larger sub-Neptunes ($\gtrsim2.4\re$) host low-mean-molecular-weight (MMW) H$_2$/He-dominated envelopes \citep[e.g.,][]{benneke_sub-neptune_2019,benneke_water_2019,mikal-evans_transmission_2020, kreidberg_tentative_2022}. 
Despite this, our understanding of the global composition of sub-Neptunes remains limited, both because of the restricted wavelength coverage offered by HST, and because of the inherent envelope mass-metallicity degeneracy in the interior composition of sub-Neptunes  \citep{rogers_three_2010,luque_density_2022,rogers_conclusive_2023}. This degeneracy consists in the same planetary mass and radius being consistent with a massive, high-molecular-weight volatile layer (e.g., H$_2$O), or a lighter H$_2$/He-dominated envelope. This becomes especially true for the less-explored planets in the vicinity of the radius valley (1.5--2.2$\,\re$), with drastically distinct implications for composition, chemical regimes, and formation histories. 

The first compositions proposed to explain sub-Neptunes, derived by extending the core-accretion predicted mass-metallicity relations from gas giants to the lowest masses, were akin to smaller versions of gas giant planets, with primordial H$_2$/He atmospheres of varying mean molecular weights and metallicities surrounding solid mantles and cores \citep{fortney_framework_2013}. Today, alternative scenarios are also being considered. On one hand, the ``water world" scenario proposes planets with ice-derived metal-rich mantles, found in their supercritical states due to the planets' high irradiation levels, and leading to highly volatile-enriched --- or even metal-dominated --- envelopes \citep{mousis_irradiated_2020,aguichine_massradius_2021,pierrehumbert_runaway_2023, piaulet_evidence_2023}. On the other hand, other models propose H$_2$/He-dominated layers living atop liquid water oceans \citep[``Hycean worlds";][]{hu_unveiling_2021, madhusudhan_habitability_2021, madhusudhan_chemical_2023} or magma surfaces \citep{kite_atmosphere_2020, schlichting_chemical_2022, zilinskas2023, shorttle_distinguishing_2024}, with a defined surface-atmosphere boundary. In general, similarly to Uranus and Neptune, it remains observationally unknown how efficient mixing is in sub-Neptune envelopes, and whether the transitions from atmosphere to mantle to core are discrete or gradual \citep{helled_uranus_2020, vazan_new_2022}.


From a planet formation and evolution perspective, the different possible sub-Neptune compositions can stem from multiple pathways. For instance, sub-Neptunes enriched in water and heavy elements are thought to form further away from the star, where large amounts of water and volatiles are efficiently accreted in the form of solid material, before migrating to their close-in position \citep{lambrecths_separating_2014, morbidelli_great_2015}. Alternatively, `dry' sub-Neptunes with smaller volatile contents are thought to form within the ice line, where volatiles are not available as ices, and where most of the mass is obtained through the accretion of drifting rocky pebbles \citep{johansen_forming_2017}. However, it was recently shown that planets that form beyond the ice line can become smaller super-Earths \citep{venturini_nature_2020}, and that sub-Neptunes dominated by high-molecular-weight volatiles (HMWVs, i.e., H$_2$O, CH$_4$, CO, CO$_2$, N$_2$, NH$_3$) can form within the ice line in multi-planet systems \citep{bitsch_dry_2021} or from magma ocean-hydrogen interactions \citep{kite_water_2021}. In all cases, it remains unclear how metal-rich the end products of each formation process are \citep{bean_nature_2021}. Hence, obtaining empirical constraints on the composition of small sub-Neptunes near the radius valley is crucial for deepening our understanding of how these planets form.


The \textit{James Webb Space Telescope} (\textit{JWST}) provides us with the opportunity to not only detect multiple species in sub-Neptune atmospheres, but also to infer their interior structure from their upper atmosphere's chemical inventories. With \textit{JWST}'s precision and wide wavelength coverage, we can precisely assess the presence or absence of the main C-, N-, O- and S-bearing molecular absorbers in the upper atmospheres of sub-Neptunes, and look for tracers of atmosphere-interior interactions \citep{yu_how_2021,hu_unveiling_2021,tsai2021inferring,schlichting_chemical_2022}. 

\begin{figure}[t!]
\begin{center}
\includegraphics[width=1.0\linewidth]{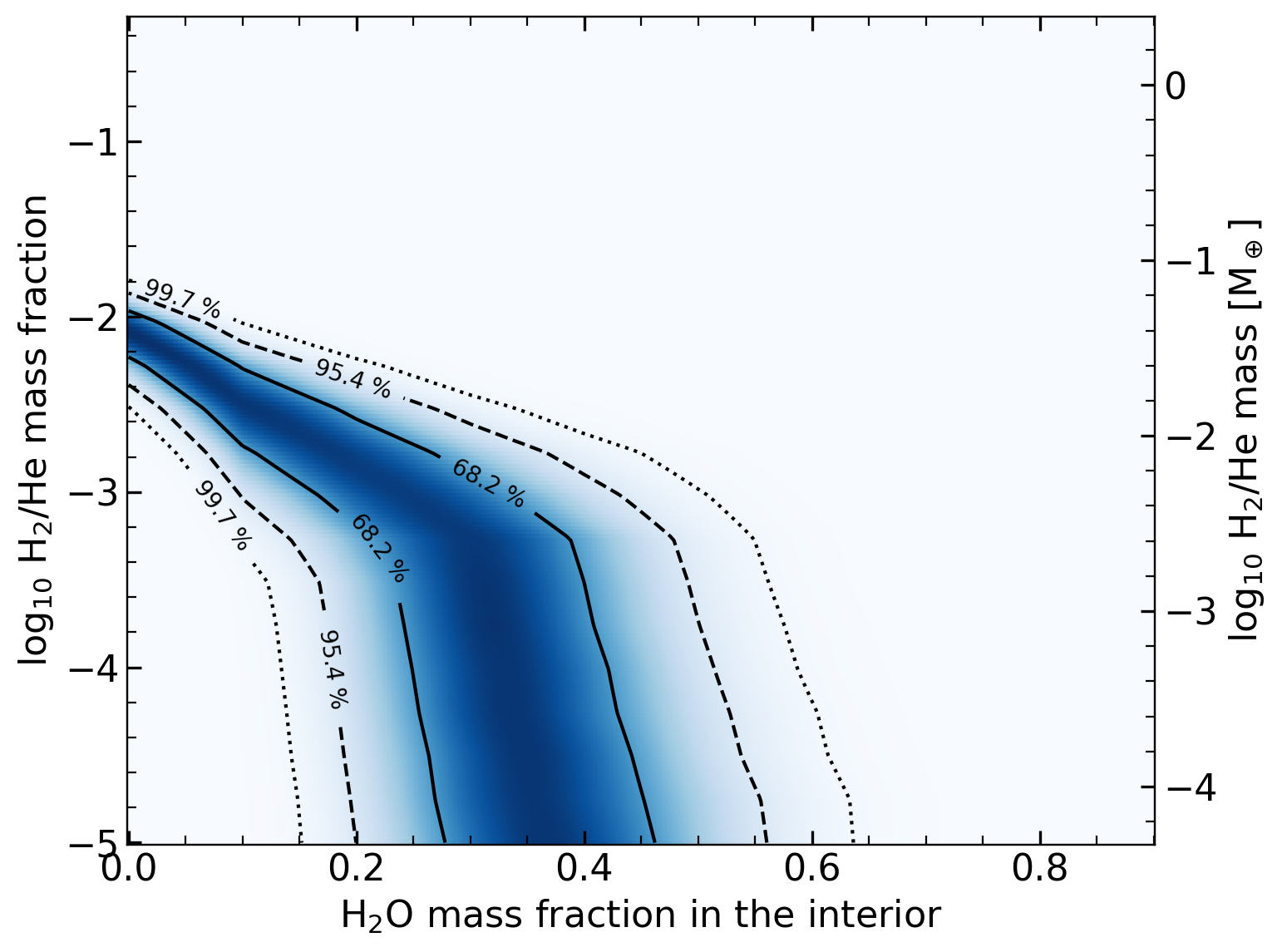}
\end{center}
\vspace{-5mm}\caption{Fit to the mass and radius of TOI-270\,d for a traditional stratified 3-layer interior structure with a H$_2$/He layer atop of a H$_2$O mantle and a rocky/iron core. Posterior probability density (blue shading) as a function of the H$_2$O mass fraction and the mass fraction of a H$_2$/He layer. The contours of 1, 2, and 3$\sigma$ confidence are outlined. Details of the interior+atmosphere model in \citet{piaulet_evidence_2023}. For this traditional stratified 3-layer interior structure, about 1\% of the planet's mass in H$_2$/He can explain the bulk density, however it can be equally well-matched with a 40\% H$_2$O mass fraction and anything in between these two end-member cases.}
\label{fig:MassRadiusFit}
\end{figure}

The transiting sub-Neptune TOI-270\,d \citep{gunther_super-earth_2019} represents an ideal target to unveil the composition of a small exoplanet with \textit{JWST} transit spectroscopy. Orbiting a M3V star, the 2.2\,R$_\oplus$ and 4.78\,M$_\oplus$ planet \citep{vaneylen_masses_2021} is found right in the degenerate space where models of stratified envelopes agree with the planet mass and radius for scenarios ranging from up to $\sim$1\% of the planet's mass in a H$_2$/He envelope to up to $\sim$40\% of the planet's mass in a deep water-dominated HMWV envelope \citep[Figure~\ref{fig:MassRadiusFit}, see also][]{rogers_three_2009, acuna_characterisation_2021, piaulet_evidence_2023}. With an equilibrium temperature of 354\,K \citep[for a 0.3 Bond albedo;][]{vaneylen_masses_2021}, TOI-270\,d is found in a slightly warmer temperature regime than K2-18\,b \citep[T$_\mathrm{eq, A_B=0.3}$=255\,K;][]{benneke_water_2019}. As we will show in this work, this temperature turns out to be in a sweet-spot, allowing TOI-270\,d's envelope to be miscible and virtually cloud-free. Importantly, TOI-270\,d's transmission spectroscopy metric \citep[TSM;][]{kempton_framework_2018} of 97 is much higher than K2-18\,b's (TSM = 42), therefore enabling a much higher signal-to-noise in the transit spectrum, and opening the door to a detailed atmosphere characterization.

The low temperature of TOI-270\,d also makes it more likely to host a clear upper atmosphere. While transit observations of multiple sub-Neptunes, such as the infamous GJ 1214\,b \citep{kreidberg_clouds_2014, kempton_reflective_2023}, led to featureless transmission spectra caused by the presence of opaque high-altitude aerosols in their atmospheres, recent trends have shown that cooler sub-Neptunes (T$_{\mathrm{eq}}\lesssim$ 400\,K) tend to display clear atmospheres with fewer clouds and hazes \citep{brande_clouds_2024}. In fact, HST transit observations of TOI-270\,d were able to successfully detect a H$_2$-rich atmosphere with an absorption feature tentatively attributed to water \citep{mikal-evans_hubble_2023}. 

Recent \textit{JWST} transit spectroscopy of the slightly cooler K2-18\,b led to the discovery of CH$_4$ and CO$_2$ in its upper atmosphere \citep{madhusudhan_carbon-bearing_2023}. Initial interpretations concluded that the presence of both species, combined with the absence of NH$_3$ which is assumed to be dissolved in a water ocean, were signs of a Hycean world nature for the planet. The presence of CO$_2$ in the upper atmosphere (not expected in chemical equilibrium) could be explained by a liquid water ocean under a thin H$_2$/He atmosphere, as CO$_2$ is the chemically-preferred carbon-bearing molecule in the presence of abundant H$_2$O \citep{hu_unveiling_2021, madhusudhan_chemical_2023, madhusudhan_carbon-bearing_2023}. However, recent studies offer alternative interpretations of the observed spectrum. For example, a silicate magma surface underneath an H$_2$ atmosphere could also explain the data at hand, with magma acting as the solvent and source for the NH$_3$ and CO$_2$ respectively \citep{shorttle_distinguishing_2024}. Moreover, climate calculations with a radiatively consistent atmosphere incorporating effects of water vapor feedback argue against the possibility of a liquid water ocean \citep{innes2023runaway,pierrehumbert2023runaway,leconte20243d}. It has also been pointed out that without a source of CH$_4$ at the bottom of the atmosphere, sustaining large amounts of CH$_4$ in a thin Hycean envelope is difficult when considering photochemical processes. Rather, a 100 $\times$ solar-metallicity envelope with quenching could reproduce the observations with more standard assumptions \citep{wogan_jwst_2024}. 

With TOI-270\,d occupying a similar low-temperature sub-Neptune regime, methane is similarly expected to be the main absorber in its atmosphere under chemical equilibrium conditions (unless the metallicity is very high). Moreover, it was recently shown that the mass and radius of TOI-270\,d can be fit with a Hycean world structure having a 200--500 km deep water ocean under a thin H$_2$/He layer \citep{rigby_ocean_2024}; however, as in the case of K2-18\,b, self-consistent climate modeling indicates that a liquid water ocean is very unlikely and that if water is present in the interior, it should be in a supercritical phase \citep{pierrehumbert_runaway_2023, innes2023runaway}. These arguments apply even more strongly to TOI-270\,d, which has a higher effective temperature.  Obtaining abundance ratios of CH$_4$, CO$_2$, H$_2$O and NH$_3$ represents a crucial step in assessing interior processes, photochemical processes, as well as vertical mixing effects in the endeavor to discern between the various proposed scenarios and constrain the composition of one of the highest-SNR sub-Neptunes near the radius valley. 

In this work, we present a 0.6--5.2~$\mu$m transmission spectrum of the 2.2~R$_\oplus$ planet TOI-270\,d, revealing multiple absorption features across the entire transmission spectrum and giving unprecedented insight into the nature of sub-Neptunes and their atmospheres. In Section 2, we present the \textit{JWST NIRISS} and \textit{NIRSpec} observations, and the data reduction performed for this study. In Section 3, we describes our atmosphere models to interpret the transmission spectrum of TOI-270\,d. Section 4 presents the results of the atmospheric analysis of the \textit{JWST/NIRISS+NIRSpec} transit spectrum, followed by a discussion of TOI-270\,d's interior structure and the metal enrichment of its envelope. Section 6 presents a sub-Neptune classification scheme in light of this work, and we discuss the atmospheric chemistry and cloudiness in Section 7. Section 8 gives a summary discussion of the nature of TOI-270\,d and puts the findings into the context of Earth and Venus as well as Uranus and Neptune.

\section{Observations and Data Analysis}\label{sec:obs_red}

Our team observed the transiting temperate exoplanet TOI-270\,d with the \textit{James Webb Space Telescope} as part of a large survey of sub-Neptunes and water-world candidates (JWST GO 4098; PI Benneke \& Evans-Soma). To cover the full wavelength range between 0.6 and 5.3 $\mu m$ for this bright star ($J$=9.1), we observed one transit with \textit{JWST NIRISS/SOSS} covering 0.6 to 2.8~$\mu m$ on Feb 6, 2024 and one transit with \textit{JWST NIRSpec/G395H} covering 2.7 to 5.2~$\mu m$ on Oct 13, 2023. This spectral range covers all major molecular features in the near-IR (H$_2$O, CO$_2$, CH$_4$, NH$_3$, CO, H$_2$S) enabling us to observationally probe for a wide variety of plausible atmospheric scenarios.

\subsection{\textit{JWST NIRISS/SOSS}}

We observed one transit of TOI-270\,d with the Single Object Slitless Spectroscopy (SOSS) mode \citep{albert2023soss} of \textit{JWST}'s \textit{NIRISS} instrument \citep{doyon2023niriss}, using the SUBSTRIP256 readout pattern which captures the full SOSS 0.6 -- 2.8\,$\mu$m waveband across two diffraction orders. The time series observations (TSO) consisted of 189 integrations with three groups per integration, yielding 16.482\,s of effective integration time, and a total exposure time of 5.3\,h. The exposure was centered on the transit of TOI-270\,d, and included 2.38\,h of baseline before and 0.78\,h after the transit. The timing of the observations was also judiciously chosen such that the target spectrum is unaffected by contamination from any background stars which can complicate spectral extraction and further analyses \citep[e.g.,][]{feinstein_early_2023, radica2023, fournier-tondreau2024}. The observations were reduced using the \texttt{NAMELESS} and \texttt{supreme-SPOON} pipelines as described below. Both reductions show good agreement between the resulting transmission spectra, with most data points agreeing at less than 0.5$\sigma$. The \texttt{NAMELESS} spectrum is used in the final atmospheric analysis.

\subsubsection{\texttt{NAMELESS} Reduction}
We reduce the NIRISS/SOSS observations of TOI-270\,d using the \texttt{NAMELESS} pipeline \citep{feinstein_early_2023,coulombe_broadband_2023}, with a few modifications which are noted below and are also described in further detail in Coulombe et al.~2024 (submitted). We begin from the uncalibrated raw data and go through the following steps of stages 1 and 2 of the STScI \texttt{jwst} reduction pipeline \citep{bushouse_2023}: super-bias subtraction, reference pixel correction, non-linearity correction, jump detection, ramp-fitting, and flat-fielding.

After these steps, we proceed with bad pixel correction where we flag pixels that consistently show null, negative, or abnormally high counts. The values of the bad pixels that have been flagged are then updated using cubic two-dimensional interpolation for all integrations. We subsequently subtract the non-uniform background by scaling independently the two regions of the STScI model background which are separated by the sharp jump in background flux, similar to the procedure presented in \cite{Lim2023}. We also correct for remaining cosmic rays that were not flagged during the \texttt{jwst} pipeline stage 1 steps by computing the running median of all individual pixels, using a window size of 10 integrations, and clipping all values that are more than 4$\sigma$ away from their running median. Moreover, we correct for the 1/$f$ noise by scaling all columns independently as described in \cite{coulombe_broadband_2023}. Because the SUBSTRIP256 mode of NIRISS/SOSS covers the three spectral orders, multiple scaling values must be considered across a single column. For the Order 1 spectral trace, the scaling factors and 1/$f$ values are computed using only pixels that are less than 30-pixel away from the center of the trace, as the amplitude of the 1/$f$ noise can vary significantly over the length of a single column. The 1/$f$ values derived from the first order are then subtracted from the full columns. We also compute scaling factors and 1/$f$ values for the spectral trace of Order 2, considering only pixels that do not overlap with Order 1, and subsequently subtract these derived 1/$f$ values from these pixels only. Finally, we extract the spectroscopic light curves from the first and second spectral orders using a simple box aperture with a diameter of 36 pixels.  

\begin{figure*}[t!]
\begin{center}
\includegraphics[width=0.4\linewidth]{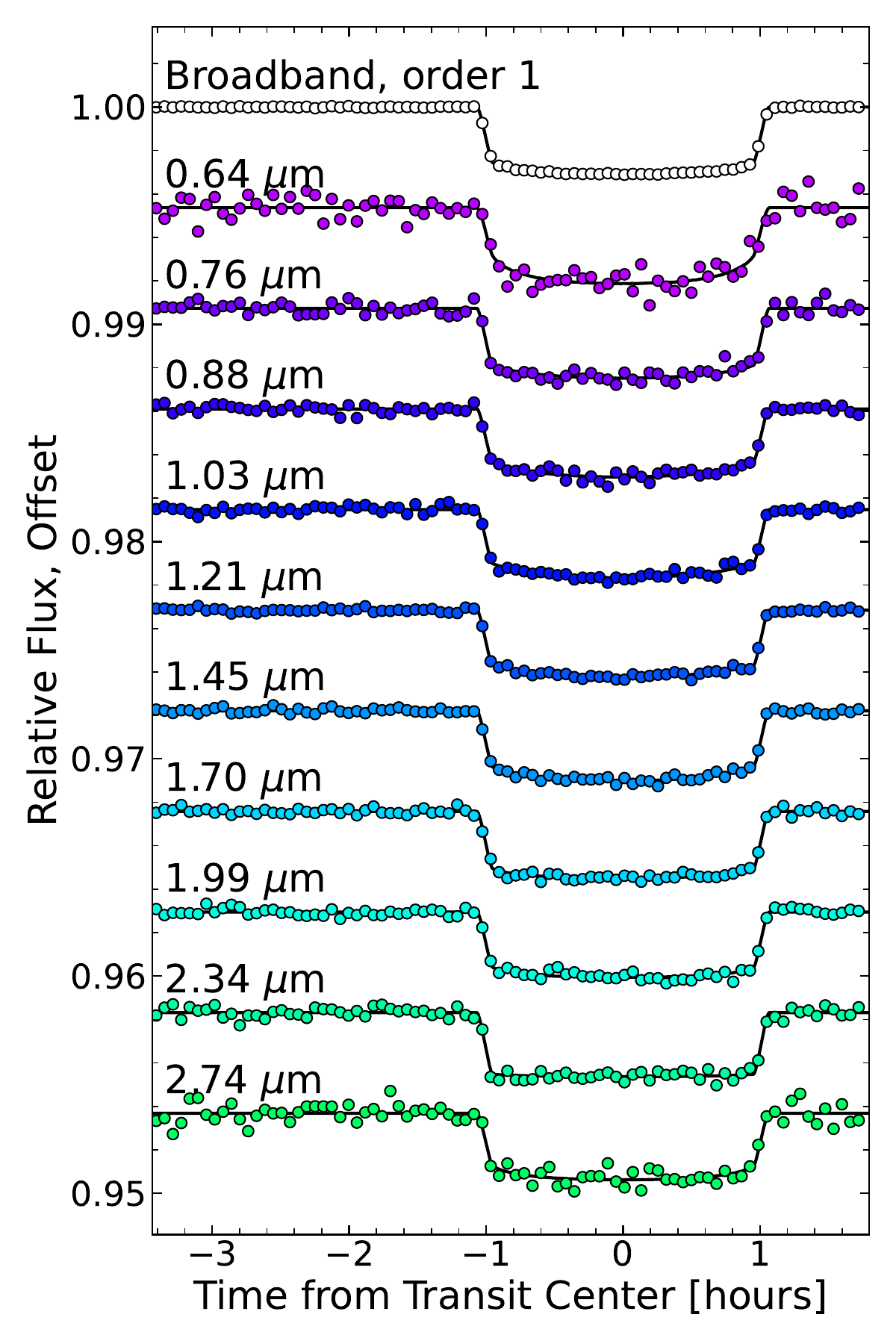}
\includegraphics[width=0.4\linewidth]{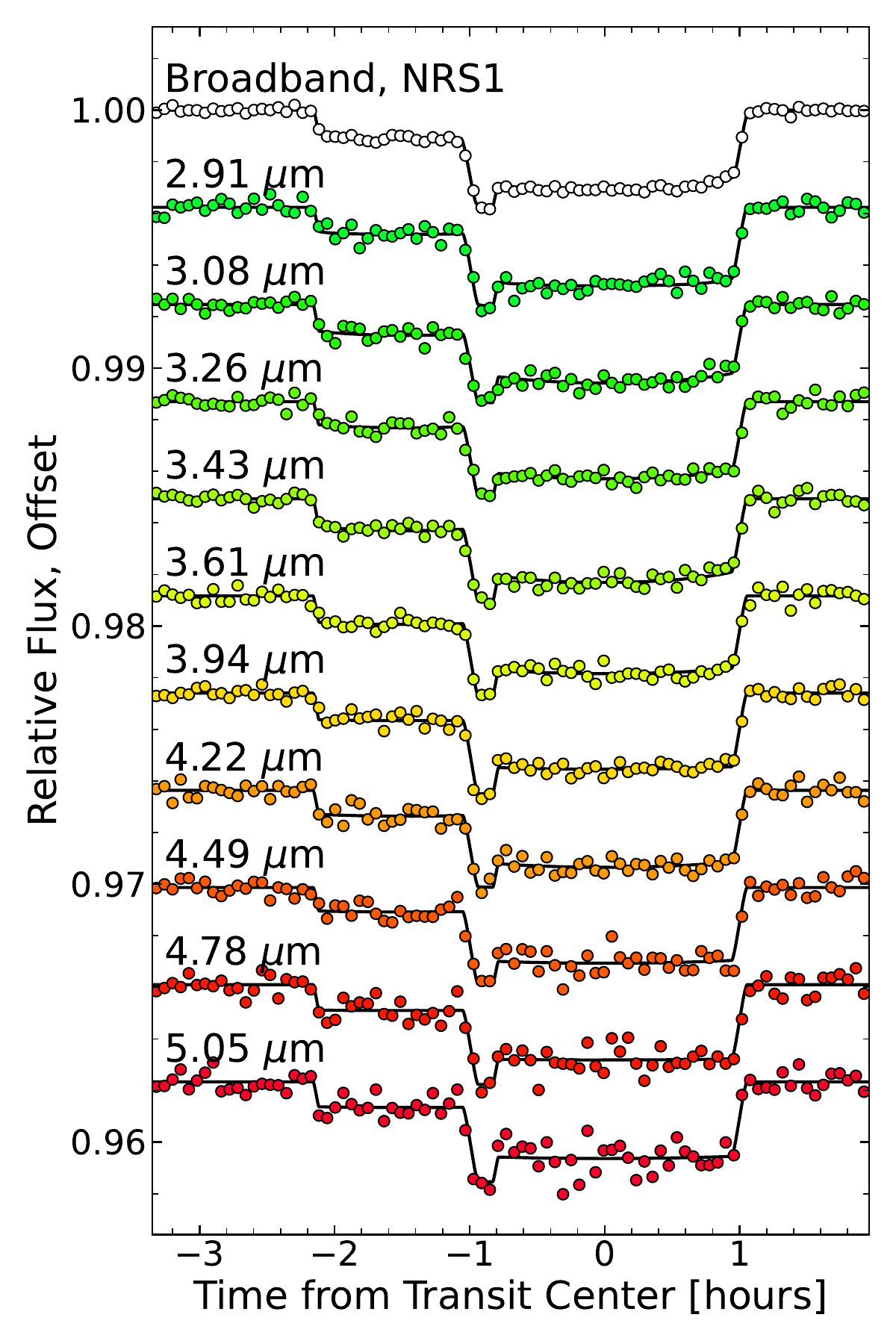}
\end{center}
\vspace{-5mm}\caption{Broadband and spectroscopic light-curve fits of the NIRISS/SOSS \textbf{(left)} and NIRSpec/G395H \textbf{(right)} transits of TOI-270\,d. For both instruments, examples of 10 normalized and systematics-corrected light curves are shown (colored points) along with their best-fitting transit models (black lines) and labelled with the central wavelength of the spectroscopic channel. Broadband light-curve fits (white) are also shown at the top for Order 1 in the NIRISS/SOSS case, and for NRS1 in the NIRSpec/G395H case. For clarity, the spectroscopic light curves are plotted with a relative flux offset. For NIRSpec/G395H, light curves with wavelengths below 3.8$\,\mu$m are from the NRS1 detector and the remaining from NRS2. For visual purposes, the light curves are also binned in time in groups of 10 (NIRISS/SOSS) and 20 (NIRSpec/G395H) data points, resulting in 3.6 minute bins. A full transit of planet b occurs at the start of the NIRSpec/G395H exposure, with both transits overlapping shortly as planet d starts transiting. All light curves are well-behaved, with no signs of spot-crossing events or H$\alpha$ flares.} 
\label{fig:lightcurves}
\end{figure*}

\subsubsection{\texttt{NAMELESS} Light Curve Fitting}
We perform the white light and spectroscopic light-curve fitting using the \texttt{ExoTEP} framework \citep{benneke_spitzer_2017,benneke_sub-neptune_2019,benneke_water_2019,roy_is_2022,roy_water_2023}. We use \texttt{ExoTEP} to simultaneously fit for the transit light curves, generated using the \texttt{batman} package \citep{kreidberg_batman:_2015}, and a systematics model in a Markov Chain Monte Carlo (MCMC) scheme \citep{foreman-mackey_emcee:_2013}. 

We begin with the analysis of the white light-curve, which is produced by summing all wavelengths of the first spectral order. The astrophysical parameters considered in the light-curve fit are the semi-major axis ($a/R_*$), the impact parameter ($b$), the mid-transit time ($T_0$), and the planet-to-star radius ratio ($R_p/R_*$), which are given as inputs to \texttt{batman}. Other parameters considered in the MCMC are the two coefficients of the quadratic limb darkening law, the photometric scatter, and a two-parameter linear systematics model. We consider wide and uniform priors for all parameters. We use 4 walkers per free parameter and run the white light-curve
fit for 10,000 steps, discarding the first 6,000 steps as burn-in. The resulting orbital parameters are found in Table~\ref{table:wlc_fit}.

We proceed with the spectroscopic light-curve fits by fixing the values of the semi-major axis, impact parameter, and mid-transit time to their best-fit values from the white light-curve fit. The light-curves have been binned at fixed resolving powers of R = 50 and R = 10 for spectral orders 1 and 2, respectively, prior to fitting. As for the white light-curve fit, we consider the planetary radius, quadratic limb-darkening coefficients, photometric scatter, and the two linear systematics model parameters. We again consider 4 walkers per free parameter and run the chains for 5,000 steps for each wavelength bin, discarding the first 3,000 steps as burn-in. The final NIRISS/SOSS transmission spectrum extracted from the light-curve fits is shown in Figure \ref{fig:TransmissionSpectrum}.

\begin{table*}
\caption{Derived orbital parameters from the fits of our TOI-270\,d \textit{JWST NIRISS/SOSS} Order 1 and NIRSpec/G395H NRS1 white light-curves extracted using \texttt{NAMELESS} and \texttt{Eureka!}, respectively. The limb-darkening coefficients given are the ($q_1$, $q_2$) values following the parametrization of \cite{kipping_efficient_2013}. We use the stellar radius ($R_*$ = $0.378\pm0.011$\,$R_\odot$) and effective temperature ($T_{\mathrm{eff}}$ = $5306\pm70$\,K) from \cite{vaneylen_masses_2021} to derive the planetary radius and equilibrium temperature values.}
\centering
\begin{tabular}{lccc}
\hline
\hline
Parameter      & NIRISS/SOSS Order 1             & NIRSpec/G395H NRS1         \\
\hline
Mid-transit time $T_0$ [BJD - 2400000.5]     &  60346.477984$_{-0.000032}^{+0.000033}$  &  60221.301725$_{-0.000084}^{+0.000087}$  \\
Planet-to-star radius ratio $R_p/R_*$   & 0.05364$_{-0.00014}^{+0.00015}$   &  $0.05379_{-0.00028}^{+0.00031}$ \\
Semi-major axis $a/R_*$   &   42.21$_{-0.82}^{+0.65}$  & 41.91$_{-1.85}^{+0.98}$\\
Impact parameter $b$      &   0.179$_{-0.116}^{+0.087}$  & 0.101$_{-0.069}^{+0.070}$\\
Limb-darkening coefficients ($q_1$, $q_2$) &  (0.093$_{-0.017}^{+0.020}$, 0.202$_{-0.068}^{+0.083}$)  & (0.093$_{-0.033}^{+0.039}$, 0.129$_{-0.093}^{+0.179}$)\\
\hline
\hline
Derived &  & \\
\hline
Planetary radius $R_p$ [$R_\oplus$] & 2.216$_{-0.064}^{+0.065}$ & 2.223$_{-0.066}^{+0.065}$ \\
Inclination $i$ [deg] & 89.76$_{-0.12}^{+0.16}$ & 89.862$_{-0.106}^{+0.096}$ \\
Equilibrium temperature $T_\mathrm{eq,A_B = 0}$ [K] & 381$\pm$8 & 384$_{-9}^{+10}$\\
Equilibrium temperature $T_\mathrm{eq,A_B = 0.3}$ [K] & 349$_{-7}^{+8}$ & 351$_{-8}^{+10}$\\
\hline
\label{table:wlc_fit}
\end{tabular}
\end{table*}

\subsubsection{\texttt{supreme-SPOON} Reduction}
We reduce the SOSS TSO using the \texttt{supreme-SPOON} pipeline \citep{feinstein_early_2023, radica2023, Radica2024}, closely following the steps laid out in \citet{Radica2024}, except for a few modifications which we note here. We perform the standard \texttt{supreme-SPOON} Stage 1 calibrations, including subtraction of the superbias, correction of non-linearity effects, cosmic ray detection, and ramp fitting. Due to the low number of groups, we flag cosmic ray hits using a sigma-clipping algorithm in the temporal domain, instead of the standard up-the-ramp flagging used in \citet{radica2023}. Loosely based on the method of \citet{nikolov_hubble_2014}, all pixels which deviate by more than five sigma from a running median in time are flagged as cosmic ray hits.

In Stage 2, we perform the background subtraction in a piece-wise manner, that is, we separately scale the standard SOSS background model on either side of the background ``step'' \citep[e.g.,][]{Lim2023, fournier-tondreau2024}. We find optimal pre- and post-step scaling values of 0.44652 and 0.46411, respectively. Moreover, using \texttt{supreme-SPOON} v1.3.0, we perform the 1/$f$ noise correction at the integration level, using the newly-implemented scale-achromatic-window method (Radica~et al.~2024b, in prep). Similar to methods described in \citet{feinstein_early_2023} and \citet{Holmberg2023}, the 1/$f$ contribution is estimated in a window of 30 pixels surrounding the target trace to better reflect the fact that the 1/$f$ noise is not constant over a single detector column. Finally, we extract the stellar spectra using a simple box aperture with a width of 30 pixels, since the expected contamination of Order 1 by Order 2 is negligible for this target \citep[e.g.,][]{Darveau-Bernier2022, Radica2022}.

\subsubsection{\texttt{supreme-SPOON} Light Curve Fitting}
We fit the extracted light curves following the procedure in \citet{Radica2024}. We first construct two white light curves by summing all the flux in Order 1, and wavelengths from 0.6 -- 0.85\,$\mu$m in Order 2. We jointly fit a \texttt{batman} transit model \citep{kreidberg_batman:_2015} to each white light curve, assuming a circular orbit and fixing the planet's orbital period to 11.3796\,d \citep{vaneylen_masses_2021}. All orbital parameters (time of mid-transit, $T_0$, impact parameter, $b$, scaled semi-major axis, $a/R_*$) are shared between orders, but the transit depths and quadratic limb-darkening parameters are fit separately. We also include an error inflation term for each order, which is added in quadrature to the extracted errors.

To handle low-amplitude correlated noise which is apparent in the light curves we fit for a slope in time, as well as linearly detrend against the first two detector principal components \citep[see e.g.,][]{coulombe_broadband_2023}, which roughly correspond to a drift in the vertical position of the target trace on the detector and a beating pattern thought to be caused by the telescope's thermal cycling \citep[e.g.,][]{coulombe_broadband_2023, McElwain2023}. After doing the above fit, we notice that some correlated noise remains in the residuals, with a power spectrum suggestive of stellar granulation noise \citep[e.g.,][]{Kallinger2014, Pereira2019, Radica2024}. Following \citet{Pereira2019}, and our recent work in \citet{Radica2024}, we account for this granulation signal using a Gaussian process (GP) with a simple harmonic oscillator kernel as implemented in \texttt{celerite} \citep{Foreman-mackey2017}. This adds an additional three parameters to the fit: the characteristic amplitude of the GP, which is fit separately to each order, and the characteristic timescale, which is shared between orders. Our final fit thus has 22 free parameters, for all of which we use wide, uninformative priors. We fit the light curves using \texttt{pymultinest} \citep{Buchner2016} as implemented in the \texttt{juliet} package \citep{espinoza_juliet_2019}.

We bin the extracted spectra to a constant resolution of $R=50$ before fitting the spectrophotometric light curves. In each bin, we fix the orbital parameters to the best-fitting values from the white light curve fits, and fit for the scaled planet radius ($R_p/R_*$), two parameters of the quadratic limb-darkening law following the \citet{kipping_efficient_2013} parameterization, as well as an error inflation term and the linear models with time and detector principal components described above. Furthermore, we account for the stellar granulation signal by linearly scaling the best-fitting white light curve GP model in each bin \citep[e.g.,][]{Radica2024}. This results in a fit with 12 free parameters per bin. Examples of several light curves and their best-fitting models are shown in Figure~\ref{fig:lightcurves}.

\begin{figure*}[t!]
\begin{center}
\includegraphics[width=0.9\linewidth]{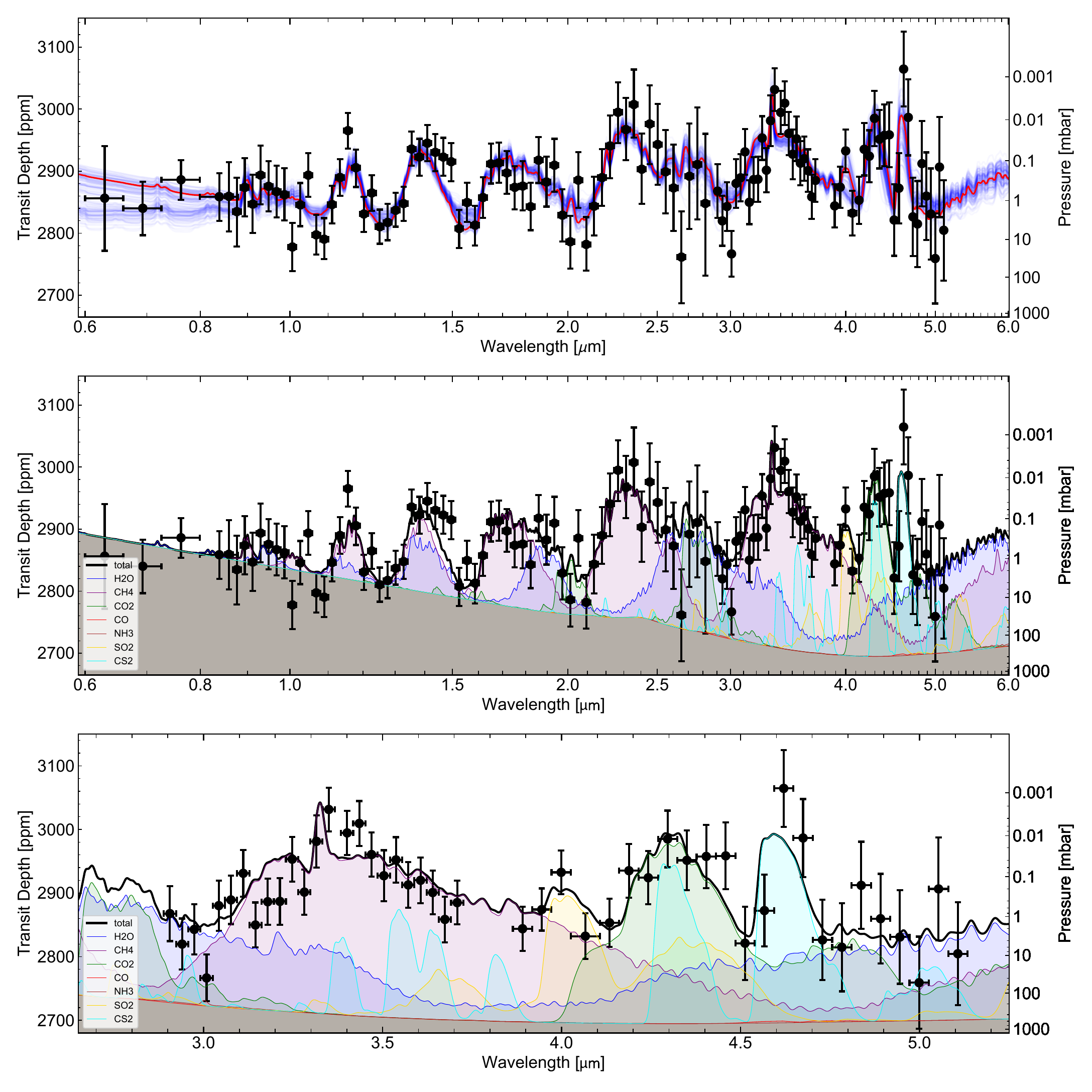}
\end{center}
\vspace{-5mm}\caption{\textit{JWST/NIRISS + NIRSpec} transmission spectrum of the temperate $2.2\,R_\oplus$ exoplanet TOI-270\,d. The top panel shows the observed transit depths (black points) compared to a random sampling of the model transmission spectra in the atmospheric retrieval posterior (blue), with the overall best-fitting model shown in red. The middle panel illustrates the contributions of individual molecular absorbers in the atmosphere of TOI-270\,d, with a zoom on the NIRSpec G395H data shown in the bottom panel. Strong vibrational bands of methane gas are detected at 1.15, 1.4, 1.7, 2.3, and 3.3~$\mu$m (purple), with H$_2$O absorption contributing as well at 1.4 and 1.9$\,\mu$m. In addition, CO$_2$ absorption is detected at 4.2--4.5~$\mu$m (green) and the increased transit depth around 4.7~$\mu$m is a possible signature of CS$_2$ (cyan). The inclusion of SO$_2$ improves the fit at 4.1~$\mu$m. The combined effect of cloud and haze opacity, which becomes significant at wavelengths below 1.1\,$\mu$m, is illustrated through the lower gray envelope of the model spectra.}
\label{fig:TransmissionSpectrum}
\end{figure*}

\subsection{\textit{JWST/NIRSpec G395H}}

We observed another transit of TOI-270\,d with the Near Infrared Spectrograph \citep[NIRSpec;][]{jackobsen_near-infrared_2022, birkmann_near-infrared_2022} using the G395H grism which covers the 2.7 -- 5.2 $\mu$m wavelength range.
Our NIRSpec Bright Object Time Series (BOTS) consisted of 1763 integrations of 11 groups. The observation covered 2 hours of pre-transit baseline, the transit itself, as well as 1 hour of after-transit baseline. The observations also captured a transit of TOI-270\,b, the innermost planet of the TOI-270 system (see Figure~\ref{fig:lightcurves}). The transit of planet b begins 1 hour before that of planet d and the two planets simultaneously occult a portion of the stellar disk for approximately 25 minutes before the end of the transit of TOI-270\,b. A thorough transmission spectroscopy analysis of planet b will be presented in a follow-up article. We used the S1600A1 (1.6\,×\,1.6) slit and the SUB2048 subarray, (two 2048\,×\,32~pixels detectors) with the NRSRAPID read-out mode to achieve an effective integration time of 10.824 s including resets, and an 83.3\% observing efficiency.  The observations were reduced using the \texttt{Eureka!}~and \texttt{Tiberius} pipelines as described below. The \texttt{Eureka!}~reduction is used in the final atmospheric analysis.

\subsubsection{\texttt{Eureka!} Reduction}
Our data reduction of the NIRSpec/G395H observations uses the \texttt{Eureka!}~framework \citep{bell_eureka_2022}, closely following the steps from the WASP-39\,b analysis \citep{alderson_early_2023}. We start our reduction from the uncalibrated data products, and use \texttt{Eureka!}'s Stage 1 to calibrate the detector images and perform the ramp fitting. \texttt{Eureka!}'s Stage 1 mainly acts as a wrapper around the standard STScI \texttt{jwst} calibration pipeline \citep{bushouse_2023}, which we follow up to the jump detection step, which we skip as it often leads to pixels being erroneously flagged as outliers \citep{alderson_early_2023, may_double_2023}. We then calibrate the integration level frames using \texttt{Eureka!}'s Stage 2, a wrapper around the standard \texttt{jwst} pipeline Stage 2. We skip the flat fielding and photometric calibrations steps as they are not needed for our transit measurements.

We use Stage 3 of \texttt{Eureka!}~to perform the background subtraction and the extraction of the time series of stellar spectra. This step starts by correcting for the curvature of the NIRSpec/G395H trace by shifting the detector columns by whole pixels, in order to bring the peak of the PSF (in the cross-dispersion direction) to the center of the subarray. This curvature correction applied to all integrations is computed from the median integration frame, and is smoothed with a running median to ensure it is robust to hot or cold pixels which are outliers even in the median frame. The pixel shifts are applied with periodic boundary conditions, meaning that the pixels moved above the subarray reappear at the bottom, ensuring all the pixels are still available for background subtraction. We perform a column-by-column background subtraction by fitting and subtracting a flat line from each column of the frames, using pixels that are further away than 6 pixels from the central row (of the curvature corrected frames) for the background fitting. In order for the background subtraction not to be biased by outliers, a double iteration of 10$\sigma$ outlier clipping is performed along the time axis, along with a 3$\sigma$ outlier clipping iteration in the spatial dimension. We finally perform an optimal extraction step, with an aperture defined by the central 9 rows of our curvature-corrected frames, and with the optimal extraction weights derived from the median frame. During Stage 3, the vertical shift of the trace from one frame to the other is measured using cross-correlation, and the average PSF width at each integration is measured by summing all columns and fitting a Gaussian to the profile.

\subsubsection{Double-Transit Light Curve Fitting}
We carry out the light-curve fitting of our \texttt{Eureka!} \textit{JWST/NIRSpec G395H} time series by simultaneously fitting for both transit light curves of TOI-270\,b and TOI-270\,d along with a joint systematics model. The astrophysical model is generated using the \texttt{batman} package \citep{kreidberg_batman:_2015} and compared to the data in a Bayesian analysis using the Markov Chain Monte Carlo sampler \texttt{emcee} \citep{foreman-mackey_emcee:_2013}. The light-curve analysis is performed by first generating and cleaning the raw light curves, then fitting the white light curve, and finally performing the spectroscopic light curve fits.

We start from the sequence of stellar spectra provided by \texttt{Eureka!}~and generate the light curves to be fitted. We discard NRS1 detector columns with wavelengths shorter than 2.86\,$\mu$m, as the throughput of the instrument becomes negligible at these short wavelengths. During this step, we define the wavelength bins of the spectroscopic light curves in order to have 50 equal-width (in wavelength) bins. Spectroscopic light curves are generated by simply co-adding the flux of all pixel-resolution light curves that fall in a given bin. We also generated a broadband white-light curve by adding the flux from every pixel of the NRS1 detector together. 

The astrophysical model is constructed by adding two \texttt{batman} transit models $T_\text{b}$ and $T_\text{d}$ for planets b and d, respectively. To correct for instrument systematic trends in the transit light curves, we simultaneously fit a two-parameter analytical linear ramp function in time along with the transit model. The full normalized light-curve model is then
\begin{equation}
f(t)= \left[ T_\text{b}(t) + T_\text{d}(t) - 1\right] \cdot \left[c + v(t-t_0)\right],
\label{eq:double_transit_mod}
\end{equation}
where $c$ and $v$ are the normalization factor and the slope of the linear systematics model, respectively. The slope is described relative to the start of the time series $t_0$.

We start the light-curve fitting by performing the analysis of the white-light curves. The main astrophysical outputs of the white-light-curve analysis are the global transit parameters ($a_\text{b}/R_\star$, $b_\text{b}$, $a_\text{d}/R_\star$, and $b_\text{d}$), the mid-transit times ($T_{0,\text{b}}$ and $T_{0,\text{d}}$), and the white-light-curve transit depths for both planets, which are all used in the \texttt{batman} light-curve modeling. Other parameters that were fitted include the photometric scatter, the coefficients of the limb-darkening, as well as the instrument systematics model parameters. The limb-darkening coefficients used follow the parametrization of \cite{kipping_efficient_2013} and the same set of coefficients is used for the astrophysical model of both planetary transits. We use wide uniform priors for all free parameters. For the MCMC, we use 4 times as many walkers as the number of free parameters in our fits and we run the broadband light-curve fit for 30,000 steps. The first 18,000 steps are discarded as burn-in when deriving median values and uncertainties from the fit. The orbital parameters derived from the MCMC fit to the NRS1 white light-curve are presented in Table~\ref{table:wlc_fit}. The systematics-corrected white-light curve and the best-fit astrophysical model are shown in Figure \ref{fig:lightcurves}.

We fix the values of the orbital parameters of planets b and d ($a/R_\star$, $b$, $T_0$) to the best-fit values found by the white-light-curve analyses and proceed to independently perform the spectroscopic light-curve fits of both the NRS1 and NRS2 detectors. In each spectroscopic bin, we thus fitted for the two transit depths, the photometric scatter, the limb darkening coefficients and the systematics model parameters. We considered the same uniform priors as for the white-light curve fit. We again used 4 walkers per parameter in our fit and the MCMC chains ran for 10,000 steps for each wavelength bin, with the first 6,000 steps discarded as burn-in.
A few systematics-corrected spectroscopic light curves are displayed in Figure \ref{fig:lightcurves}. The final NIRSpec/G395H transmission spectrum of TOI-270\,d using the \texttt{Eureka!} reduction and custom light curve fitting described above is shown in Figure \ref{fig:TransmissionSpectrum} along with the NIRISS/SOSS data. We note that we discard three bins ($\lambda$ = 3.38, 4.07, and 5.15\,$\mu$m) that are significant outliers in $R_p/R_*$ or show abnormally high uncertainties.

\subsubsection{\texttt{Tiberius} Reduction}
We perform an independent data reduction of the NIRSpec/G395H observations using the \texttt{Tiberius} pipeline \citep{Kirk2017Tiberius, Kirk2021Tiberius}. Similarly to the \texttt{Eureka!}~reduction we start with the uncalibrated data products and run individual steps from Stage 1 and 2 of the \texttt{jwst} pipeline, skipping the jump detection step as well as flat fielding and photometric calibrations, but before the ramp-fitting step we apply a median column-by-column background correction at group-level which is shown to reduce 1/f noise systematics in NIRSpec/G395H observations \citep[e.g.][]{alderson_early_2023}. 

Following closely the steps from previous \textit{JWST} data reductions with \texttt{Tiberius} \citep[e.g.][]{lustig-yaeger_jwst_2023, Kirk2024GJ341}, we extract the time-series spectra by fitting a Gaussian to each column of the frame and using a polynomial to fit the trace centers, where we used a running median and clipped $5\sigma$ outliers to improve the trace-fitting. Additional column-by-column background subtraction is performed using a linear fit with a mask that includes the trace and an additional 3 pixels on each side. The flux is extracted using an aperture full width of 8 pixels.

\begin{figure}
    \centering
    \includegraphics[width=0.45\textwidth]{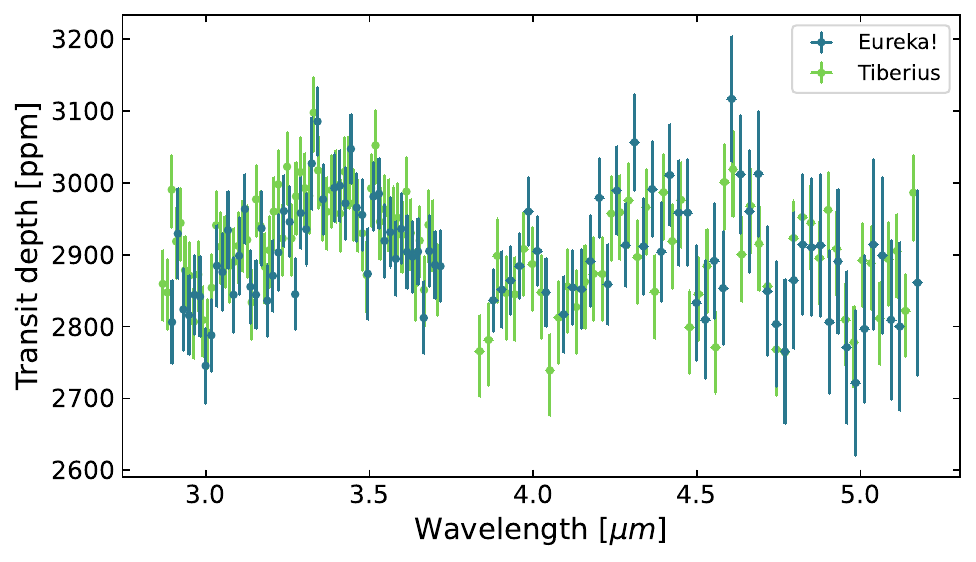}
    \vspace{5mm}
    \vspace{-5mm}\caption{\textit{JWST NIRSpec/G395H} transmission spectrum of TOI-270\,d obtained using two independent reduction and light curve fitting pipelines, \texttt{Eureka!}~and \texttt{Tiberius}, described in detail in the text. Both spectra are in agreement across the spectral range, showing all the same features.}
    \label{fig:G395H_transmission_spectrum}
\end{figure}


\subsubsection{\texttt{Tiberius} Light Curve Fitting}

We perform the light-curve fitting of our \texttt{Tiberius} extracted time series considering the same astrophysical and systematics model as Equation \ref{eq:double_transit_mod}. We carry out the fitting using an adapted \texttt{Tiberius} light-curve fitting code \citep{Ahrer2022tiberius_polychord} that employs the \texttt{PolyChord} nested sampling algorithm \citep{Handley2015_polychordI,Handley2015polychordII}. The white-light curve is generated by summing the flux over the whole G395H wavelength range and the spectroscopic light curves are produced using bins of 20 pixels. Similar to the light curve fitting performed on the \texttt{Eureka!}~reduction, we fit for the orbital parameters of both planets and consider a single set of limb-darkening coefficients. We consider the standard quadratic parametrization for the limb-darkening coefficients and fix the parameter $u_2$ to values generated using the \texttt{LDTK} package \citep{parviainen_exoplanet_2015}. For the nested sampling fit, we used 100 times as many live points as the number of free parameters.

The orbital parameters of both planets are fixed to their white-light curve best-fit value at the spectroscopic light-curve fitting stage. We fit each wavelength bin independently and consider the same astrophysical and systematics models as for the white-light curve, using again 100 times as many live points as the number of free parameters for the nested sampling. The transmission spectra of TOI-270\,d extracted using the two independent pipelines and light curve fitting codes are shown in Figure\,\ref{fig:G395H_transmission_spectrum} and present good agreement over the full G395H wavelength range.

\section{Atmosphere Models}\label{sec:retrieval}

\subsection{\texttt{SCARLET} Free-Chemistry Atmospheric Retrieval}

We perform atmosphere retrievals on our TOI-270\,d transmission spectrum using the SCARLET framework \citep{benneke_atmospheric_2012,benneke_how_2013,knutson_featureless_2014,kreidberg_clouds_2014,benneke_strict_2015,benneke_sub-neptune_2019, benneke_water_2019, pelletier_where_2021, pelletier_vanadium_2023, roy_is_2022, roy_water_2023, coulombe_broadband_2023}. SCARLET parameterizes the molecular abundances, the cloud deck pressure, the hazes Rayleigh slope, and the upper atmosphere temperature to fit our transmission spectrum. The framework uses a Bayesian nested sampling analysis \citep[with single ellipsoid sampling;][]{skilling_nested_2004, skilling_nested_2006} to obtain the posterior probability distribution of our atmospheric parameters and the Bayesian evidence of our models. 

During the nested sampling exploration, SCARLET produces a forward atmosphere model for each set of parameters. At each model evaluation, SCARLET produces a 40-layer one-dimensional atmosphere in hydrostatic equilibrium, computes the molecular opacities, and then computes the transmission spectrum for that atmosphere by solving the radiative transfer equation. The model spectrum, initially produced at R = 16,000, is then binned to the resolution of our observed spectrum so that the likelihood evaluation can be performed. For our free chemistry retrievals, we consider CH$_4$, CO$_2$, H$_2$O, CO, NH$_3$, CS$_2$ and SO$_2$ as the parameterized molecules, while H$_2$ and He fill up the remainder of the atmosphere \citep{benneke_how_2013}. 

Our SCARLET free chemistry retrieval assumes that the molecular abundances are well-mixed and thus remain constant with altitude. Each molecule has a log-uniform Bayesian prior extending from 10$^{-10}$ to 1 in volume mixing ratio. The temperature is also assumed to be constant with altitude, and the prior on that parameter extends uniformly from 50 -- 600\,K. The cloud opacity is modeled by a cloud deck top pressure ($p_\mathrm{Cloud}$), which blocks all incoming light, and by a haze scattering parameter ($c_\mathrm{Haze}$), which multiplies the usual Rayleigh scattering coefficients of our atmosphere models in order to simulate excess Rayleigh scattering from photochemical hazes and molecules not parametrized in the retrieval. The priors on both cloud parameters are log-uniform and extend from 10$^{-7}$ bar to 10$^2$ bar for the cloud top pressure, and from 10$^{-6}$ to 10$^5$ for the Rayleigh slope multiplicative factor. In total, our full atmosphere retrieval includes 10 free parameters (see Table \ref{tab:detect}).

For each molecule, we infer the significance of detection in Table~\ref{tab:detect} using the Bayesian model comparison approach introduced in \citet{benneke_how_2013}, which involves removing one molecule at a time and then determining the Bayes factor between the retrieval model with all molecules (reference model) and the retrieval model with that one molecule removed.

\subsection{Quenched-chemistry Atmospheric Retrieval}

From a formation and interior structure perspective, the elemental composition of the deep atmosphere of sub-Neptunes is of great interest, often more so than the detailed chemistry of their upper atmosphere. To address this, we introduce a novel quenched-chemistry atmospheric retrieval setup that enables us to directly retrieve constraints on the elemental carbon-to-hydrogen (C/H) and oxygen-to-hydrogen (O/H) ratios of the deep atmosphere (1--10~bar) while taking into account that the vertical temperature structure of the deep atmosphere is not known a priori and poorly constrained by the data.

The underlying idea is that complex chemical kinetics, photochemistry, and diffusion models consistently predict that the photospheres of warm exoplanets (1-100~mbar) are generally in a chemically quenched state. Chemical reactions are slower towards the higher layers of the mid-atmosphere, where temperatures are colder. Thus, the presence of vertical mixing introduces a transition from the thermochemical-equilibrium regime of the deep atmosphere to a mixing-driven ``quench-chemical" regime at higher altitudes \citep{visscher_quenching_2011,fortney_beyond_2020, hu2021photochemistry}. As a result, the molecular abundances of a large part of the photosphere are set by thermochemical-equilibrium at the quench point, i.e. the point where the vertical mixing rate overcomes the chemical reaction rates. The quench-chemical regime at higher altitudes is therefore set almost exclusively by the temperature and pressure of the quench point, in combination with the overall elemental abundances in the deep atmosphere.

Motivated by this, we introduce a new parameterization that describes the molecular composition in the photosphere with only four parameters including the elemental oxygen-to-hydrogen ratio [O/H], the elemental carbon-to-hydrogen ratio [C/H], the temperature at the quenching point ($T_\mathrm{quench}$), and the pressure at the quenching point ($p_\mathrm{quench}$). As priors, we choose a log-uniform prior on C/H and O/H between $10^{-6}$ and 1, a uniform prior on $T_\mathrm{quench}$ between 400 and 1200~K, as well as a log-uniform prior on $p_\mathrm{quench}$ between 0.1~bar and 10~bar. The quenched-chemistry atmospheric retrieval also includes free parameters for the cloud-top pressure, haze opacity, and the photospheric temperature, $T_\mathrm{phot}$, with the same priors as in the case of the free-chemistry retrieval. While the quenching temperature sets the molecular abundances at the quenching pressure, the photospheric temperature sets the hydrostatic equilibrium as well as the molecular opacities at the photosphere (the pressures probed by the observations). Equivalently to [C/H] and [O/H], we also run retrievals with a log-uniform priors on the overall metallicity [M/H] ($0.1-10000$~x~solar) and the carbon-to-oxygen ratio (C/O) ($10^{-3}-10^3$). We obtain identical results to the C/H and O/H parameterization for this highly constraining \textit{JWST/NIRISS+NIRSpec} data set.


\subsection{EPACRIS Photochemistry-Thermochemistry Models}\label{sec:EPACRIS_method}
We constructed a detailed chemical reaction network for highly-metal (O, C, N, S)-rich atmospheres at the temperatures and pressures relevant for TOI-270\,d using the Reaction Mechanism Generator (RMG) \citep{Gao_2016, RMG-database}, and incorporated the automatically generated network to the chemical-kinetics module of the ExoPlanet Atmospheric Chemistry \& Radiative Interaction Simulator \citep[EPACRIS,][]{yang2024automated}. The RMG-generated reaction network, which addresses thermochemical equilibrium and quenching, is combined with an existing network tailored for photochemistry in low-temperature planetary environments \citep{hu2021photochemistry}, with the updates in \cite{wogan_jwst_2024} as well as the addition of CS$_2$-, CS-, and OCS-related photochemical reactions \citep{tsai_VULCAN_2017}. 
The final network included 99 species and 2019 reactions. 

To simulate the atmosphere of TOI-270\,d, we calculated the pressure-temperature profiles under radiative-convective equilibrium for the elemental abundances in Table~\ref{tab:detect} and Bond albedo values of 0.3, 0.7, and 0.9, using the climate module EPACRIS (Scheucher et al. in prep.). The climate module solves the radiative fluxes using the 2-stream formulation of \cite{heng_radiative_2018} and performs the moist adiabatic adjustment using the formulation of \cite{graham2021multispecies}. Water is considered as a condensable in these models. Using the chemical network and pressure-temperature profiles, self-consistent photochemical kinetic-transport atmospheric models were generated for a pressure-independent eddy diffusivity ranging in $10^4-10^6$ cm$^{2}$\,s$^{-1}$. In general, the overall atmospheric chemistry of TOI-270\,d is dominated by disequilibrium chemistry, and we find that the resulting abundances of CH$_4$, CO$_2$, and H$_2$O at the pressures probed by the transmission spectrum are consistent with the requirements from the planet's spectrum as summarized in Table~\ref{tab:detect}. See Section~\ref{sec:atmos_chem} for a discussion of the atmospheric chemistry results.

\subsection{Photochem Photochemistry-Thermochemistry Model}
\label{sec:Photochem_method}
To test the robustness of our photochemical-climate calculations, we also used the model Photochem \citep{wogan_jwst_2024,wogan2023origin} to simulate atmospheric speciation in TOI-270\,d. The model used for TOI-270\,d is identical to the gas dwarf previously developed for K2-18\,b \citep{wogan_jwst_2024} except that planet bulk properties, bolometric insolation, and input metallicity and C/O ratios have been adjusted. We assumed the bulk atmospheric metallicity and C/O ratio retrieved from the transmission spectrum, a well-mixed deep atmosphere (K$_{zz}$ = 10$^8$ cm$^2$\,s$^{-1}$), and a Jupiter-like upper atmosphere eddy-diffusion profile \citep[i.e., troposphere K$_{zz}$ = 10$^3$-10$^4$ cm$^2$\,s$^{-1}$ and increasing with altitude following][]{hu2021photochemistry}. We found that the self-consistent models can reproduce the observed molecular abundances (see Section~\ref{sec:atmos_chem}).

\subsection{The Dual-Grey General Circulation Model}\label{sec:gcm}
To assess the possibility of longitudinal temperature variations, we construct a simplified general circulation model (GCM) of TOI-270\,d using the Met Office’s Unified Model \citep{wood_2014, mayne_2014} with dual-grey radiative transfer using the ``super-sol” opacities taken from \citet{menou_2012}.  To maintain numerical stability, a vertical sponge with $\eta_s=0.9$ is used and a 2nd order filter is applied to the horizontal velocities in the longitudinal direction with a strength of $K=0.55$ \citep[the details of the dissipation scheme can be found in ][]{mayne_2014}. The configuration is otherwise similar to that used in \citet{christie_2022} with the substitution of clear-sky, dual-grey radiative transfer and planetary parameters appropriate for TOI-270\,d.  The atmospheric pressures range from 200 bar at the base of the computational domain to $P < 10^{-5}\,\mathrm{bar}$ at the top.    As the rotational period of TOI-270\,d is unknown, the planet is assumed to be tidally locked, and the model is run for 1000 Earth days.

\section{Atmospheric Results}\label{sec:resu}

The observed transit spectrum of TOI-270\,d prominently displays five strong vibrational bands of CH$_4$ at 1.15, 1.4, 1.7, 2.3, and 3.3~$\mu$m detected at 9.4$\sigma$ (Figure \ref{fig:TransmissionSpectrum}). In addition, the signature of CO$_2$ is clearly detected at 4.2--4.5~$\mu$m (4.8$\sigma$), and the retrieval analysis infers the presence of H$_2$O (2.5$\sigma$) as well as the potential signature of carbon disulfide (CS$_2$) at 4.6$\,\mu m$ (2.6$\sigma$). To this date, our observations of TOI-270\,d present the highest SNR transmission spectrum for any planet in the sub-Neptune regime, providing unprecedented insight into the chemistry and elemental abundances of a sub-Neptune exoplanet.

\subsection{Mean Molecular Weight (MMW)}

Transmission spectra are highly sensitive to the mean molecular weight of the atmosphere through its effect on the atmospheric scale height \citep{miller-ricci_atmospheric_2009}. This is particularly true for our combined \textit{JWST NIRISS + G395H} spectrum of TOI-270\,d because multiple detected molecular bands from the same molecules (here: the five bands of CH$_4$) can uniquely constrain the mean molecular weight \citep{benneke_atmospheric_2012,benneke_how_2013}. Intriguingly, for TOI-270\,d, we overall find a mean molecular weight of $\mathrm{MMW}=5.47_{-1.14}^{+1.25}\,\mathrm{amu}$, significantly elevated compared to the hydrogen-dominated atmospheres of giant planets (Figure \ref{fig:retrieval}). At the most basic level, this high mean molecular weight means that the atmosphere must be highly metal-enriched in molecular species heavy compared to H$_2$ and He. As we will show in the following subsections, the high mean molecular weight is the result of high molecular abundances of CH$_4$, H$_2$O, CO$_2$, and potentially CO in the atmosphere of TOI-270\,d. The atmosphere is highly enriched in carbon and oxygen compared to primordial H$_2$/He atmospheres. 

\begin{table}
\caption{\label{tab:detect} Atmospheric retrieval results for TOI-270\,d. Free composition retrieval above the line. Chemically-consistent retrieval with quenching at the bottom. The uncertainties represent the 68\% confidence intervals for all two-sided constraints and the 95\% upper limits for molecules not detected.  The detection significance values are derived from the Bayes factor following the relation in \cite{benneke_how_2013}.}
\centering
\begin{tabular}{lccc}
\hline
\hline
Parameter & Estimate  & Significance \\
\hline
         \logX{CH4} & $-1.64_{-0.36}^{+0.38}$ & $9.43\sigma$\\ 
         \logX{CO2} & $-1.67_{-0.60}^{+0.40}$ & $4.80\sigma$\\ 
         \logX{H2O} & $-1.10_{-0.92}^{+0.31}$ & $2.54\sigma$\\ 
         \logX{CO}  & $<-1.46$ \\ 
         \logX{NH3} & $<-4.27$ \\ 
         \logX{SO2} & $-4.39_{-3.33}^{+1.01}$ \\          
         \logX{CS2} & $-3.44_{-0.67}^{+0.66}$ & $2.55\sigma$\\ 
         T$_\mathrm{phot}$ [K] & $385.3_{-41.8}^{+44.2}$\\ 
       $\log c_\mathrm{Haze}$ & $<-8.26$ \\ 
   $\log p_\mathrm{Cloud}$ [bar] & $>-2.99$ \\ 
          MMW [amu] & $5.47_{-1.14}^{+1.25}$\\
$\logXratio{CO2}{CH4}$ & $-0.08_{-0.49}^{+0.41}$\\ 
$\logXratio{H2O}{CH4}$ & $0.52_{-1.10}^{+0.54}$\\ 
$\logXratio{CO}{CH4}$  & $<+0.21$\\ 
$\logXratio{NH3}{CH4}$ & $<-2.59$\\ 
\hline
  O/H by number & $0.114_{-0.040}^{+0.051}$\\ 
    O/H by mass & $0.91_{-0.32}^{+0.41}$\\ 
  C/H by number & $0.050_{-0.023}^{+0.030}$\\ 
    C/H by mass & $0.30_{-0.14}^{+0.18}$\\ 
(C+O)/H by mass & $1.23_{-0.44}^{+0.51}$\\ 
            C/O & $0.47_{-0.19}^{+0.16}$\\ 
            Metallicity [x solar] & $224.6_{-86.1}^{+98.1}$\\ 
            \Zatm\ [\%] & $58_{-12}^{+8}$\\ 
     log p$_\mathrm{quench}$ [bar] & $-0.02_{-0.64}^{+0.66}$ & \\ 
        T$_\mathrm{quench}$ [K] & $567.4_{-31.0}^{+36.7}$\\ 
     T$_\mathrm{phot}$ [K] & $371.5_{-37.9}^{+40.6}$\\ 
       $\log c_\mathrm{Haze}$ & $<-8.50$ & \\ 
        $\log p_\mathrm{Cloud}$ [bar] & $>-2.78$ & \\ 
\hline
\hline
\end{tabular}
\end{table}

\subsection{Molecular Abundances}

\begin{figure*}
\begin{center}
\includegraphics[width=1.0\linewidth]{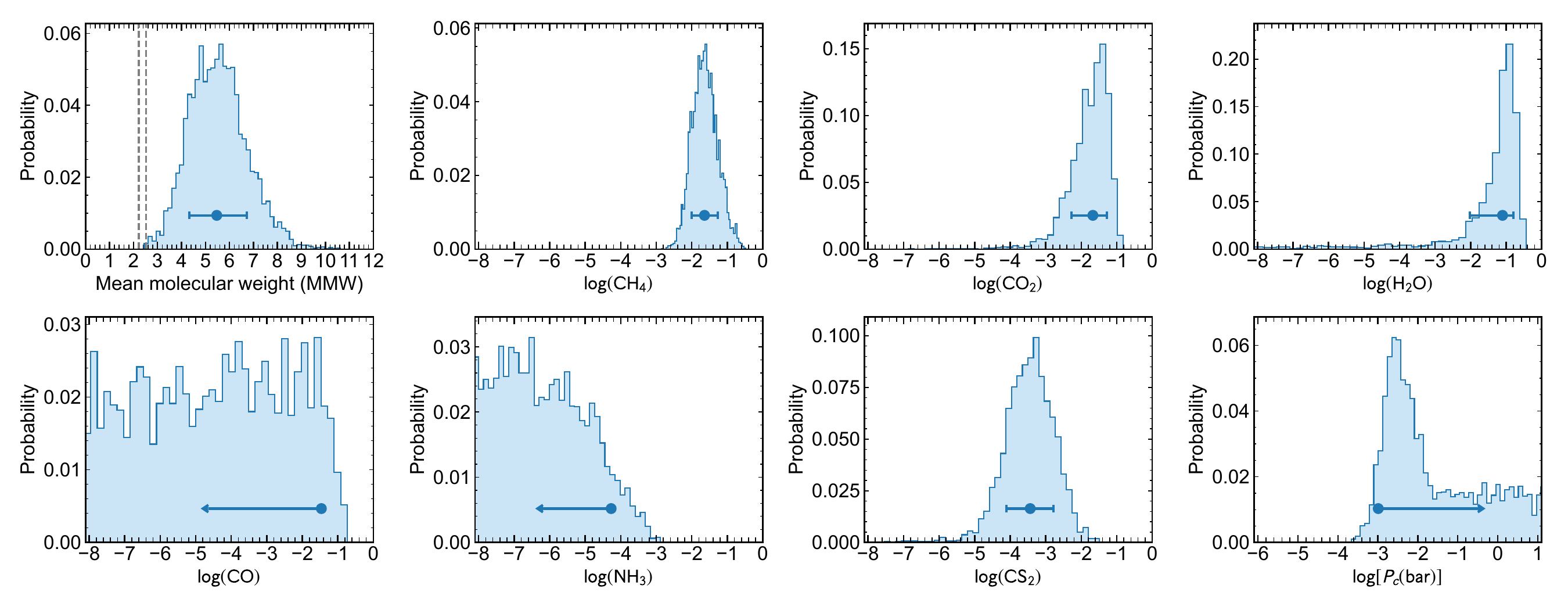}
\end{center}
\vspace{-5mm}\caption{Posterior probability distributions of the free parameters used in the SCARLET free-chemistry retrieval. The panels show the marginalized probability distributions. Vertical dashed line in the top left panel indicate the mean molecular weight of Jupiter's and Neptune's atmosphere in the Solar System.} 
\label{fig:retrieval}
\end{figure*}

Our free-chemistry retrieval analysis also provides important constraints on the individual abundances of the major carbon, nitrogen, oxygen, and sulfur-bearing molecules in the atmosphere of TOI-270\,d. We derive a log volume mixing ratio of $\logX{CH4}=-1.64_{-0.36}^{+0.38}$, making it the most precisely constrained molecule in the sub-Neptune regime to date (Figure \ref{fig:retrieval}). 
This corresponds to the atmosphere of TOI-270\,d being $2.3_{-1.3}^{+3.2}\%$ methane by number.
CO$_2$ is also present in a comparable abundance to CH$_4$, with a log volume mixing ratio of $\logX{CO2}=-1.67_{-0.60}^{+0.40}$ or $2.1_{-1.6}^{+3.2}\%$ by number. CO$_2$ is readily detected thanks to its strong vibrational band at 4.3\mum, which does not overlap with strong CH$_4$ absorption bands. The relative strength of the 4.3$\,\mu$m CO$_2$ absorption feature, when compared to the CH$_4$ features, provides a direct measure of the relative abundance of CO$_2$ and CH$_4$ resulting in a robust constraint on the CO$_2$-to-CH$_4$ abundance ratio, $\logXratio{CO2}{CH4}=-0.08_{-0.49}^{+0.41}$.

We find water vapor to also be highly abundant in the atmosphere of TOI-270\,d with a log volume mixing ratio of $\logX{H_2O}=-1.10_{-0.92}^{+0.31}$, potentially the largest component of the atmosphere by mass. Despite the high H$_2$O abundance, the water vapor is inferred only at $2.5\sigma$ in the atmosphere of TOI-270\,d because the high opacities of CH$_4$ at near-infrared wavelengths strongly overlap with those of H$_2$O (Figure~\ref{fig:TransmissionSpectrum}). As a result, H$_2$O is largely invisible in the G395H data and also mostly blocked by the strong bands of CH$_4$ in the SOSS data. Water vapor does, however, introduce a noticeable effect on the transmission spectrum of TOI-270\,d, especially near 1.9 and $3.0\mum$ as well as redwards of $4.8\mum$, from which the retrieval identifies its presence. We obtain a bounded abundance constraint for carbon-disulfide (CS$_2$) giving a log volume mixing ratio of $\logX{CS2}=-3.44_{-0.67}^{+0.66}$, which the atmospheric chemistry further discussed in Section~\ref{sec:atmos_chem}. Notably, we also obtain a strongly informative 2-$\sigma$ upper limit on the abundance of ammonia (NH$_3$) in the atmosphere of TOI-270\,d at $\logX{NH3}=-4.27$ indicating an NH$_3$ abundance below 54 parts-per-million in the photosphere at approximately 1~mbar.

\begin{figure*}
\begin{center}
\includegraphics[width=0.70\linewidth]{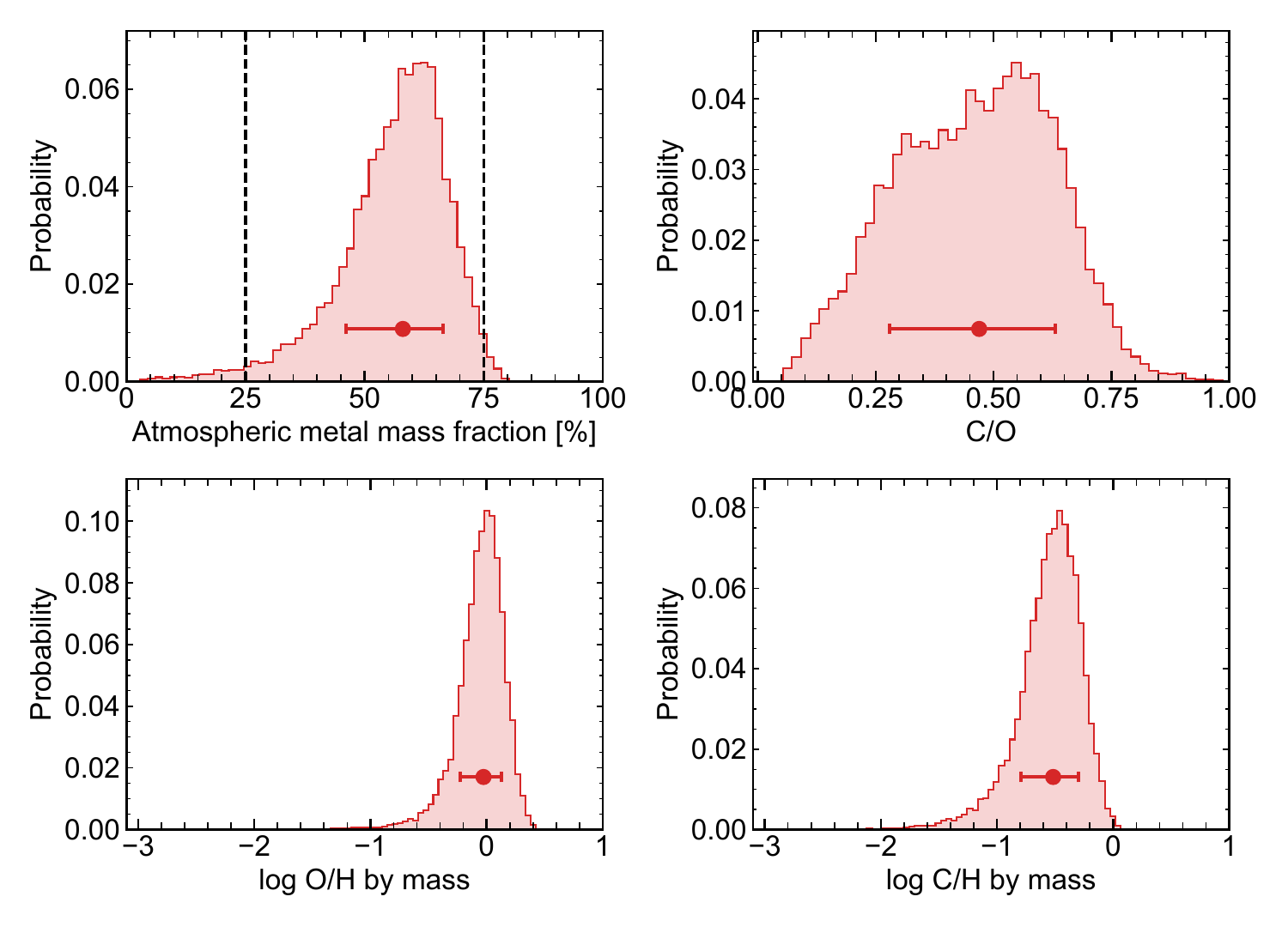}
\end{center}
\vspace{-5mm}\caption{Posterior probability distributions of the atmospheric metal mass fraction, the C/O ratio, as well as the oxygen-to-hydrogen and carbon-to-hydrogen mass ratios in the atmosphere. 1D distributions are shown marginalized over the freely parameterized temperature and pressure at the quench point in the atmosphere, the cloud and haze properties, as well as the photospheric temperature. The \Zatm and C/O posterior distributions are derived from the fitted C/H and O/H, and they are identical when fitting for metallicity and C/O instead.} 
\label{fig:Zatm}
\end{figure*}

\subsection{Elemental Abundances C/H, O/H and C/O}

Our quenched-chemistry retrieval enables us to directly retrieve the elemental abundances in TOI-270\,d's atmosphere from its \textit{JWST} transmission spectrum while taking into account the a-priori poorly constrained pressure and temperature at the quenching point (Section \ref{sec:retrieval}). We find an atmosphere highly enriched in carbon and oxygen relative to solar, consistent with the high mean molecular weight and high abundances of volatile molecules (CH$_4$, CO$_2$, H$_2$O) measured in the free-chemistry retrieval. 

Quantitatively, we derive an O/H ratio of $0.114_{-0.040}^{+0.051}$ by number, corresponding to $0.91_{-0.32}^{+0.41}$ by mass. This indicates that the total mass in oxygen atoms in the atmospheric gas is approximately the same as the mass in hydrogen atoms in the outer envelope (Figure~\ref{fig:Zatm}). Equivalently, the C/H ratio is $0.050_{-0.023}^{+0.030}$ by number and $0.30_{-0.14}^{+0.18}$ by mass. Overall, the (C+O)/H by mass is $1.23_{-0.44}^{+0.51}$, which corresponds to a $224.6_{-86.1}^{+98.1}$ fold enrichment of volatile elements compared to the proto-solar abundances. 

We find a C/O ratio in the atmosphere of TOI-270\,d of $0.47_{-0.19}^{+0.16}$. It is most likely a coincidence that our measured carbon-to-oxygen ratio is consistent with the solar value of 0.59 \citep{asplund_chemical_2021} because most of the carbon and oxygen in the atmosphere were likely accreted as ices or released as volatile molecules (CH$_4$, CO$_2$, H$_2$O) through magma-ocean/envelope interaction (Section~\ref{VMO}). Moreover, some of the oxygen is likely bound in refractory molecules that are condensed out of the photosphere. For these reasons, the measured C/O could have been widely different from the solar value. Large amounts of both oxygen and carbon were added to the atmosphere throughout its formation.

The above-mentioned constraints on C/H, O/H, and C/O are the 68\% Bayesian credible intervals after marginalizing over the a-priori poorly constrained pressure and temperature at the quench points in the lower atmosphere. Here, we parameterize $p_\mathrm{quench}$ and $T_\mathrm{quench}$ as free parameters, making no a-priori assumptions on the vertical temperature pressure profile. We find that the molecular abundances above the quench point are highly sensitive to $T_\mathrm{quench}$ but with virtually no direct dependency on $p_\mathrm{quench}$. As a result, $p_\mathrm{quench}$ is largely unconstrained by the observations, while we derive a quench point temperature of $T_\mathrm{quench}=567.4_{-31.0}^{+36.7}$\,K. Not surprisingly, this is significantly warmer than the photospheric temperature of $371.5_{-37.9}^{+40.6}$\,K; however, comparison with self-consistent radiative-convective models (Figure \ref{fig:Kzz_comparison_updated}) suggests that the quench point is relatively shallow in the mid-atmosphere below the photosphere. If $T_\mathrm{quench}$ was significantly higher, the relative strengths of CH$_4$, CO$_2$, H$_2$O and CO absorption bands could not be matched, resulting in a worse fit to the data overall. Informed by the free retrieval, we did not include the opacities of NH$_3$ in the quenched-chemistry retrieval.

\subsection{Atmospheric Metal Mass Fraction (\Zatm)}
The inferred metallicity of $224.6_{-86.1}^{+98.1}$ times solar and the C/O of $0.47_{-0.19}^{+0.16}$ translate into an atmospheric metal mass fraction of $\Zatm=58_{-12}^{+8}\%$, indicating that around half the mass of TOI-270\,d's photospheric gas is in metals rather than H and He. The transmission spectrum of TOI-270\,d enables us to directly detect carbon, oxygen, and sulfur-bearing molecular species, representing three of the four most abundant volatile-forming metals in the Universe. While the effect of the other volatile metals (N forming N$_2$, Ar, etc.) cannot individually be measured, the overall effect of their masses still increases the mean molecular weight of the atmosphere, which is well constrained by the atmospheric retrieval. The \Zatm\ thus characterizes well the total percentage of the atmospheric mass that is in volatile metals (C, O, N, S, P, as well as the noble gases) and not in H and He. It is well constrained even if the nitrogen and phosphorus mass fractions cannot be measured individually.

It is worth noting that the inferred \Zatm, C/H, O/H, and C/O from our retrieval analysis correspond to the values found in the optically thin atmosphere. Refractory elements such Fe, Mg, and Si, if present, are condensed out of the uppermost layers and do not affect the measured mean molecular weight. If they are present in appreciable amounts in the deeper layers of TOI-270\,d's envelope, then these refractory elements would further increase the metal mass fraction in the lower layers of the envelope, even if the envelope is otherwise well-mixed in the super-critical phase. This might be the case in particular if magma-ocean/envelope geochemistry \citep{schlichting_chemical_2022,misener_atmospheres_2023} transported refractories into the envelope. We therefore must distinguish between the \Zatm\ of the observable atmosphere and the overall envelope metal mass fraction (\Zenv). In general, the \Zenv\ will be higher than the \Zatm, especially because an atmospheric layer residing atop a distinct layer of lower density would result in Rayleigh-Taylor instabilities, rendering the atmosphere unstable. Furthermore, oxygen can be bound with the refractory elements into silicates at deeper layers, decreasing the O/H of the observable atmosphere. The \Zenv\ is not part of our atmospheric results analysis and must be modeled assuming scenarios for the overall volatile-to-refractory abundance ratio.

\subsection{Photospheric Temperature and Clouds \& Hazes}\label{sec:temp_clouds}
We find consistent constraints on the photospheric temperature and clouds/hazes parameters between the free-chemistry and the quenched-chemistry retrieval analysis. The retrieved photospheric temperature is $T_\mathrm{phot} = 385.3_{-41.8}^{+44.2}$\,K for the free-chemistry retrieval and $T_\mathrm{phot} = 371.5_{-37.9}^{+40.6}$\,K for the quenched-chemistry retrieval, respectively. This represents the temperature at a pressure of approximately 0.1--1\,mbar near the day-night terminator. The photospheric temperature is consistent with the equilibrium temperature assuming full heat redistribution, which is the expected circulation regime for TOI-270\,d given its slow rotation rate \citep{Pierrehumbert2019wtg}, and reasonable Bond albedo values of 0.0 to 0.3. The photospheric temperature also explains why large amounts of water vapor are readily present in the photosphere, as it is too hot to allow for water condensation (see Figure \ref{fig:gcm_pt}).

The opacities of clouds and hazes in the photosphere of TOI-270\,d, if present at all, are low even to grazing light beams throughout most of the infrared. As a result, the infrared transmission spectrum of TOI-270\,d appears largely cloud-free with only the shortest wavelengths in the NIRISS/SOSS observations ($0.6-1.1\mum$) affected by cloud and haze opacities. In our parameterization, both small-particle diffuse Rayleigh hazes and a thick gray upper cloud deck are possible, and we find that at least one of them is needed at 3.35$\sigma$. However, because the effect of cloud and haze opacities on the transmission spectrum is small for TOI-270\,d, and only the regions with the lowest molecular opacities shortwards of $\sim1.1\mum$ are affected, we cannot infer the wavelength dependence of the cloud/haze opacities and find that both small-particle hazes and a low-lying thick cloud deck are consistent with data. The constraints are consistent between the free-chemistry and the quenched-chemistry retrieval analysis. Quantitatively, we derive $p_\mathrm{cloud} > 1.0\,$mbar as the 95\% lower limit on the cloud top pressure, indicating that the atmosphere is cloud free down to at least 1~mbar (Figure~\ref{fig:TransmissionSpectrum}). The transmission spectrum does not probe deeper because the abundances of the main infrared absorbers CH$_4$, CO$_2$, and H$_2$O are hundreds of times higher than they would be for a primordial atmosphere of a gas giant.

\subsection{No Substantial Offset Between NIRISS and NIRSpec}
We performed the atmospheric retrievals both with and without an additional free parameter to account for the possibility of a systematic offset between the NIRISS/SOSS and NIRSpec/G395H data. In the former case, we find a value of $12_{-17}^{+16}$\,ppm for the offset, consistent with zero at 1$\sigma$. The atmospheric retrieval results are consistent between retrievals with and without the additional free offset parameter.

\section{Discussion of TOI-270d's Interior Structure and the Envelope's Metal Enrichment}\label{sec:implications}

\subsection{A metal-rich miscible envelope on TOI-270\,d}

The discovery that approximately half of TOI-270\,d's outer envelope's mass is in volatile metals, rather than H$_2$/He is in disagreement with the typically assumed layered interior structure models of sub-Neptunes, in which a distinct H$_2$/He layer resides on top of a metal-rich ``icy" mantle or a rock/iron core \citep{rogers_three_2009,valencia_bulk_2013}. Instead, our observations suggest an alternative scenario for sub-Neptunes like TOI-270\,d, characterized by a metal-rich miscible envelope in which H$_2$/He does not have a distinct layer at the top but is well-mixed with the high-molecular-weight volatiles (mostly H$_2$O, CO, CH$_4$) in a miscible supercritical metal-rich envelope.

In this Section, we investigate the overall volatile metal mass of TOI-270\,d and the origin of the volatiles. We introduce a matching sub-Neptune classification scheme in Section \ref{sec:classification}. It is highly plausible that TOI-270\,d presents an archetype of planets in the sub-Neptune regimes, especially the small medium-density ones near $2\,\re$, and that our metal-rich envelope scenario applies to the thousands of sub-Neptunes discovered to date.

\subsection{Miscibility of H$_2$/He and H$_2$O on Warm Sub-Neptunes}

Hydrogen and water are miscible in the gas phase as well as in the supercritical phases \citep{soubiran_miscibility_2015, pierrehumbert2023runaway,innes2023runaway}. Hydrogen, water, and all the other high-molecular-weight volatiles (HMWV), e.g., CH$_4$, CO, CO$_2$, will therefore naturally tend towards being in a well-mixed state, with temperature-dependent chemistry controlling the balance between CH$_4$, CO, and CO$_2$, but not fundamentally introducing vertical gradients in O/H as is the case for Neptune and Uranus. Neptune and Uranus host cold tropospheres resulting in the condensation and cold-trapping of H$_2$O removing it from the upper layers of the atmospheres amenable to spectroscopic characterization. 

Our measurement of a metal-rich, medium-MMW atmosphere on TOI-270\,d support this theoretical prediction of a miscible gas envelope \citep{pierrehumbert2023runaway,innes2023runaway}. As TOI-270\,d is sufficiently warm for the temperature pressure profile to cross directly from the gas phase into the supercritical phases, hydrogen and water remain miscible throughout the envelope (see also Section~\ref{sec:classification} for details).  

Refractory elements, if present, would still be locked in the hotter layers of the envelope and must have taken some oxygen with them \citep{schlichting_chemical_2022,chachan_breaking_2023}; however, the bulk of the volatile metals (C, N, O, S, Ne, etc.) could still remain largely well-mixed all the way to the observable photosphere. 





\subsection{Rock/Iron Core Mass and Total Amount of Metals in TOI-270\,d's Envelope}\label{sec:Rock_Iron_Core_Mass}

Depending on the formation location, formation models predict that sub-Neptunes can form either wet, by accreting large amounts of ices beyond the water ice line (ice-rich formation); or relatively dry,  akin to rocky planets interior to the water ice line that subsequently accrete hydrogen (dry formation) \citep{lee_breeding_2016, bitsch_dry_2021}. At the most basic level, we therefore need to distinguish between sub-Neptunes with rock-dominated interior and sub-Neptunes with volatile-dominated interiors.

Measurement of the planetary radius and mass alone cannot distinguish these formation scenarios because of the inherent degeneracy in the mass-radius diagram. Planets with a large water-dominated interior and a small hydrogen envelope can overall lead to the same bulk density as a smaller denser rock-dominated interior with a big hydrogen envelope \citep[e.g.,][]{rogers_three_2009}. For fully miscible envelopes, however, the atmospheric metal mass fraction (\Zatm) becomes the critical third measurable that together with the measured planet mass and radius enables the constraint of the individual mass fractions of the three planet components (rock/iron core, H$_2$/He in the envelope, and volatile metals in the envelope).  

\begin{figure}
\begin{center}
\includegraphics[width=1.0\linewidth]{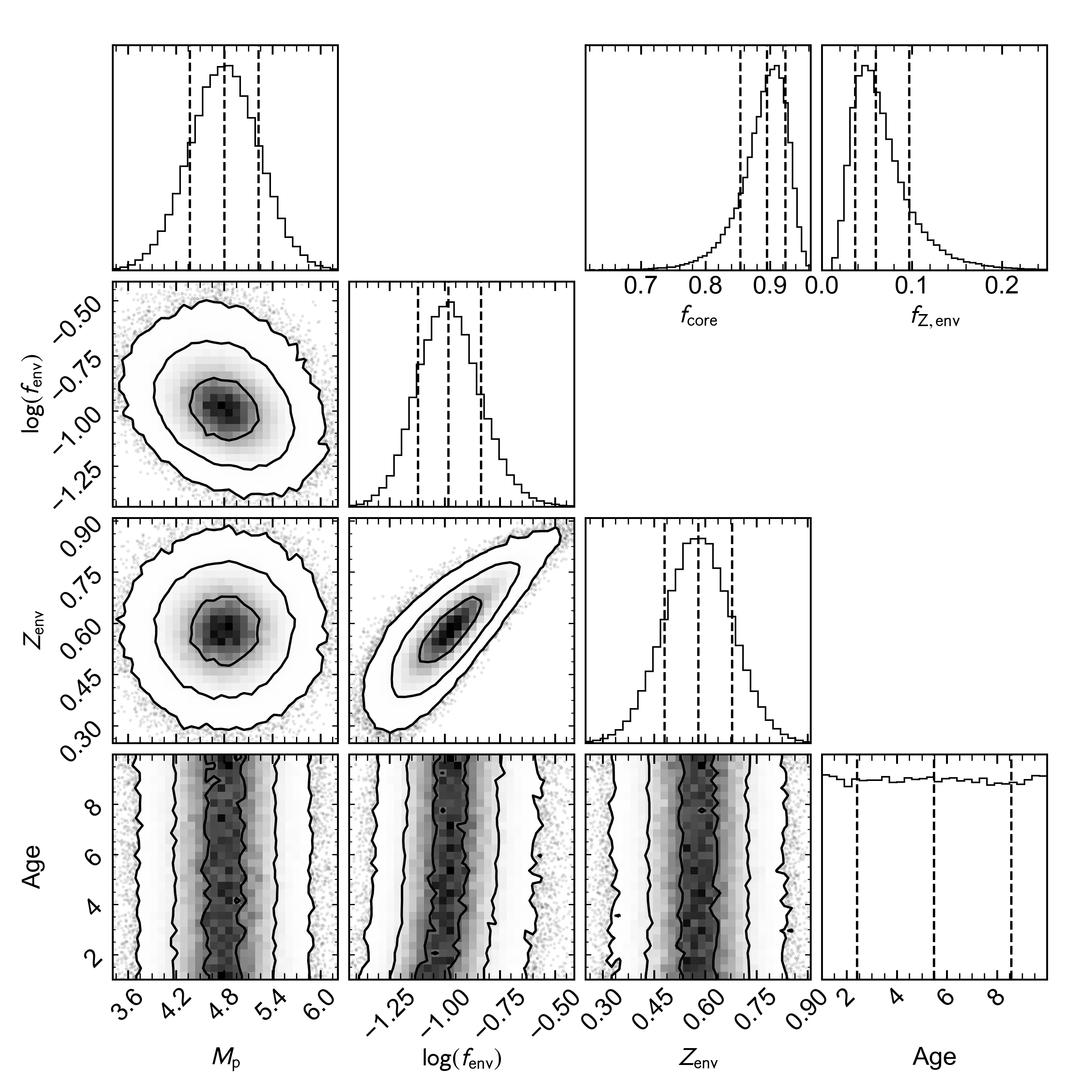}
\end{center}
\vspace{-5mm}\caption{Interior model fitting of TOI-270\,d's planetary mass and radius under consideration of the measured \Zatm\ of the miscible envelope. The marginalized posterior distributions of the planet mass $M_\mathrm{p}$, the envelope mass fraction $\log(f_c)$, the envelope metal mass fraction $Z_\mathrm{env}$, and the age of the system are shown in the panels on the diagonal, along with the 2D marginalized distribution below the diagonal. The mass fraction of TOI-270\,d's rock/iron core $f_\mathrm{core}$ is found to be $90_{-4}^{+3}$\,wt\,\% as shown in the upper-right corner together with the fraction of the planet's total mass in envelope metals ($f_\mathrm{Z,env}$).}
\label{fig:coremassfraction}
\end{figure}

For TOI-270\,d, we constrain the interior composition using models of exoplanet structure and thermal evolution \citep{thorngren_massmetallicity_2016,thorngren._connecting_2018}, in which we consider a fully miscible H$_2$/He+H$_2$O envelope on top of a rock/iron core. As prior information, we take into account TOI-270\,d's measured planetary mass ($M_\mathrm{p}=4.78_{-0.43}^{+0.43}$ \,$M_\oplus$) and a envelope metal mass fraction ($Z_\mathrm{env}$) equal to the measured atmospheric metal mass fraction ($\Zatm=58_{-12}^{+8}\%$). We then impose a log-uniform prior for the envelope mass fraction $f_\mathrm{env}$ between $10^{-4}$ and 1, and a uniform prior for the planet's age between 1~Gyr and 10~Gyr. Using MCMC, we then perform a Bayesian analysis fitting for TOI-270\,d's observed radius of $2.22\pm0.06$\,$R_\oplus$. The resulting posterior distribution is shown in Figure~\ref{fig:coremassfraction}. We infer that $90_{-4}^{+3}$\% of TOI-270\,d's total mass are in the rock/iron core, indicating a highly rock/iron-dominated interior (Table~\ref{tab:massfractions}). The remaining mass is divided into $4.3_{-0.8}^{+0.9}$\% of H$_2$/He and $6.0_{-2.3}^{+3.7}$ of volatile metals (C, N, O, S, etc.) in the envelope. The total amount of mass in volatile metals (i.e., forming H$_2$O, CH$_4$, CO, etc.) is $0.29_{-0.11}^{+0.17}\,\me$, corresponding to $1244_{-464}^{+761}$ times the mass of Earth's oceans.

\begin{table}
\vspace{-5mm}\caption{\label{tab:massfractions} Results from interior model fitting under consideration of \Zatm.}
\centering
\begin{tabular}{lccc}
\hline
\hline
Parameter & Estimate \\
\hline
Mass fraction of rock/Fe core, $f_\mathrm{core}$ & $0.90_{-0.04}^{+0.03}$\\ 
Mass fraction of total envelope, $f_\mathrm{env}$ & $0.10_{-0.03}^{+0.04}$\\ 
Mass fraction in H/He, $f_\mathrm{H/He}$ & $0.043_{-0.008}^{+0.009}$\\ 
Mass fraction in envelope metals, $f_\mathrm{Z,env}$ & $0.060_{-0.023}^{+0.037}$\\ 
Core mass, $M_\mathrm{env}$ & $4.28_{-0.47}^{+0.46}\,\me$ \\ 
Envelope mass, $M_\mathrm{env}$ & $0.50_{-0.13}^{+0.19}\,\me$\\ 
Mass in H/He, $M_\mathrm{H/He}$ & $0.207_{-0.034}^{+0.037}\,\me$\\ 
Mass in envelope metals, $M_\mathrm{Z,env}$ & $0.29_{-0.11}^{+0.17}\,\me$\\
$M_\mathrm{Z,env} / M_\mathrm{Earth\,ocean}$ & $1244_{-464}^{+761}$\\
\hline
\hline
\end{tabular}
\end{table} 
\subsection{Origin of Metal Enrichment in TOI-270\,d's Envelope}

In the following three subsections, we discuss three mechanisms to deliver the inferred volatile metals to TOI-270\,d's envelope, including magma-ocean geochemistry, ice accretion, and metal enrichment via the preferential loss of hydrogen.

\subsubsection{Metal Enrichment via Magma-Ocean Geochemistry}\label{VMO}

One efficient source of volatile metals is the geochemical creation of high-molecular-weight volatiles (e.g., H$_2$O, CH$_4$, CO) at the bottom of the envelope via its reaction with the hot magma-ocean \citep{kite_water_2021,schlichting_chemical_2022}. We performed a range of global chemical equilibrium calculations to assess the range of possible volatile concentrations as a result of hydrogen-magma ocean interactions. The calculations build on previous work \citep[for details see][]{schlichting_chemical_2022}, with the addition of C in all phases (i.e., metal, silicate and envelope). Assuming a rocky core and hydrogen-dominated envelope yields a hydrogen mass-fraction of 1--2\% and magma-ocean temperatures of about 4000--5000\,K for TOI-270\,d's mass and radius, when allowing for ages from 1--8 Gyr (see also Figure \ref{fig:MassRadiusFit}). For these parameters we find mean molecular weights (MMW) for the upper atmosphere ranging from about 3--8 with atmospheric metal mass fractions (\Zatm) ranging from about 35--80\%, which is consistent with the atmosphere retrieval results. Although a detailed one-to-one comparison between the various chemical species and their abundances is beyond the scope of this work and will be addressed in a future paper, the main species that we expect in the upper atmosphere as a result of a magma-ocean interactions with a hydrogen rich envelope are CH$_4$, H$_2$O and CO/CO$_2$.

\begin{figure*}[t!]
\begin{center}
\includegraphics[width=1.0\linewidth]{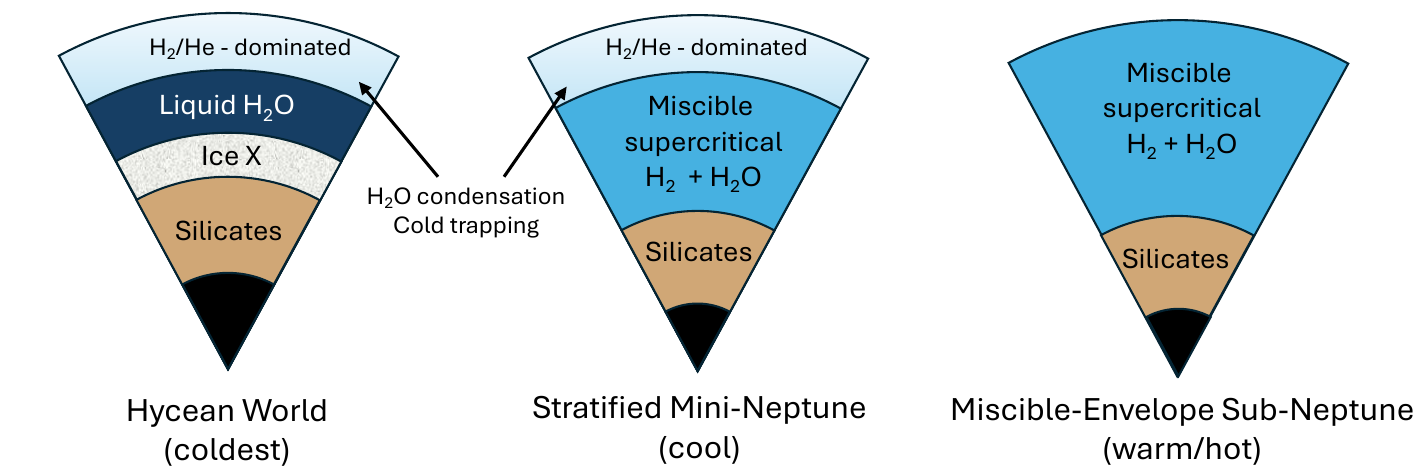}
\end{center}
\vspace{-5mm}\caption{Temperature-dependent interior structure of sub-Neptunes driven by the phase changes of H$_2$O. TOI-270\,d's high atmospheric metal mass fraction indicates that high-molecule-weight volatiles (H$_2$O, CH$_4$, CO, CO$_2$) are well-mixed with the H$_2$/He in a warm miscible envelope (right scenario). The C/H and O/H of the atmosphere is therefore much more representative of the overall envelope composition than for stratified mini-Neptune and hycean worlds.}
\label{fig:Temp_vs_InteriorStructure}
\end{figure*}

\subsubsection{Metal Enrichment via Ice Accretion}

Volatile delivery via ice accretion presents another way of increasing the metal budget of a sub-Neptune. Assuming that TOI-270\,d formed further away from its star than its current orbital position, e.g., beyond the ice line, provides a way to explain the metal enrichment of this planet. In these cooler conditions, high-molecular-weight volatiles (H$_2$O, CH$_4$, CO, etc.) are available as solids and can be efficiently accreted onto the planet in the early stages of its formation \citep{lambrecths_separating_2014, morbidelli_great_2015}. The planet would thus have assembled its metal content, and most of its mass, away from the star, and would have subsequently migrated to its close-in orbit without losing these heavy elements. A migration history for the TOI-270 system is plausible given the near-resonant orbits of the three planets \citep{terquem_migration_2007, cresswell_three_2008, ramos_planetary_2017}.

One caveat of the migration scenario in explaining the ice content of TOI-270\,d, is that it could easily have led to a steam world composition for the planet, with an even larger atmospheric metal content. Formation of planetary cores beyond the ice line can easily lead to the formation of cores with as much water as rock by mass (M$_\mathrm{H_2O} \in [0.5 \, M_\mathrm{rock}, \ M_\mathrm{rock}]$), and even after photoevaporative evolution, these planets can retain this ratio \citep{venturini_nature_2020}. This prediction is difficult to reconcile with TOI-270\,d's mass and radius, as the H$_2$ atmosphere observed on the planet allows for at most a few percent of water by mass in its interior (Figure \ref{fig:MassRadiusFit}). Hence, in order for the migration model to explain the formation of TOI-270\,d, it would imply that the efficiency of the ice accretion beyond the ice line was altered either in time (a late formation case) or by the availability of ices (e.g., a formation at the ice line with migration rapidly setting in). Another possibility, if large amounts of ices (and potentially more H$_2$/He gas) were accreted initially, is that it was subsequently lost due to efficient atmospheric escape (see below).

\subsubsection{Metal Enrichment via Hydrogen Escape}

While TOI-270\,d may have accreted its hydrogen-dominated envelope from the protoplanetary disk, the present-day enrichment in volatile species will not necessarily reflect the composition of the accreted primordial gas. The assumption from most small planet evaporation models \citep[e.g.,][]{lopez_role_2013,owen_kepler_2013,owen_evaporation_2017} that the planet's envelope mass fraction decreases over time while retaining the same composition has been challenged by the consideration of element fractionation \citep{hu_helium_2015} and preferential loss of hydrogen \citep{chen_evolutionary_2016} relative to helium and other heavier metals due to their diffusive separation. Subsequent modeling of the interior, thermal evolution and atmospheric mass-loss of small planets including this effect \citep{malsky_coupled_2020} revealed that the preferential loss of hydrogen leads to significant enrichments in the helium and metal mass fractions over billions of years, with final atmospheric metal mass fractions exceeding 10\% for the thin envelopes of planets near the radius valley \citep{gu_deuterium_2023, malsky_helium_2023}. While metals can be dragged along with the hydrodynamic hydrogen outflow, dampening the metal enhancement effect, models exploring how the loss of heavier species is impacted by the relative impact of downward diffusion from gravity and upward drag \citep{cherubim_strong_2024,louca_metallicity_2023} predict that the atmospheric compositions of planets with thin hydrogen atmospheres remain affected by mass fractionation, especially for planets with low insolation levels. 

Model predictions suggest that TOI-270\,d lies at the mean equilibrium temperature (370~K) for planets where mass fractionation can lead to an enhancement in the relative abundance of deuterium, helium and heavier metal species, especially under the impact of core-powered mass-loss \citep{cherubim_strong_2024}. The higher escape fluxes predicted for EUV-flux-driven photoevaporation would result in lower metal enhancements. Although further detailed modeling would be required to adequately assess the relative impact of the preferential loss of hydrogen and of the photochemical breakup of water \citep{bolmont_water_2017} and methane on the final atmospheric mass fractions, an order-of-magnitude tenfold increase in the atmospheric metal mass fraction from hydrogen escape remains a plausible explanation for the present-day highly supersolar metallicity of TOI-270\,d.

\section{Sub-Neptune Classification}\label{sec:classification}

To this date, the bulk compositions, interior structures, and formation histories of planets in the sub-Neptune regime remain poorly understood, even at the most basic level. Clearly, the atmospheric metal mass fraction of TOI-270\,d ($\Zatm = 58_{-12}^{+8}\%$) deviates from most existing fiducial interior models that consider a distinct hydrogen envelope on top of an ice mantle or rocky/iron core \citep[e.g, ][]{rogers_three_2009,valencia_composition_2010}. Motivated by this metal-rich atmosphere on TOI-270\,d, we revisit the interior structure of sub-Neptunes based on their observable upper envelope compositions and their interior structures. We use the term ``sub-Neptune'' as a general term for planets with radii and masses between those of Earth and Neptune, and then introduce classes of sub-Neptunes.

\begin{figure*}[t!]
\begin{center}
\includegraphics[width=0.60\linewidth]{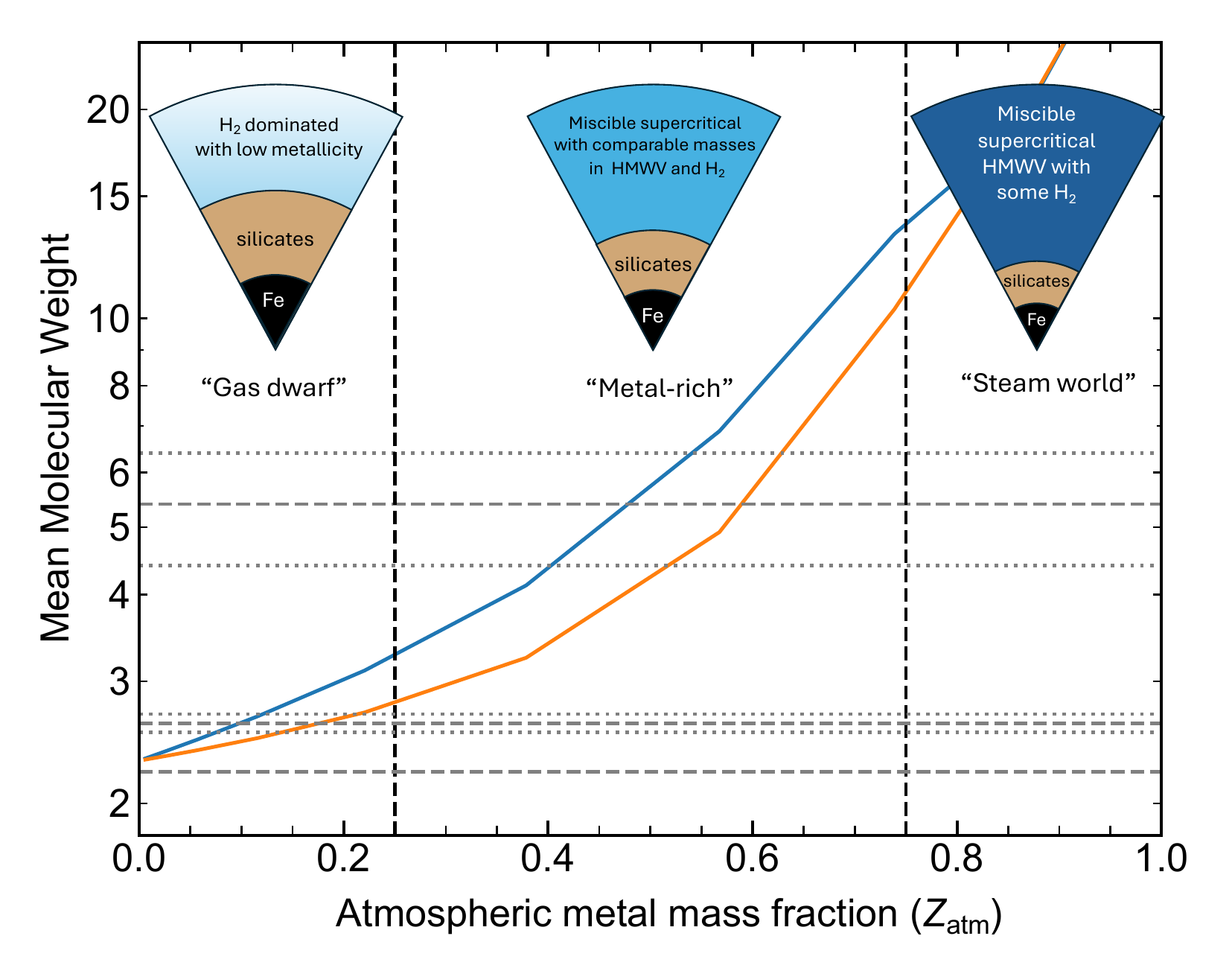}
\end{center}
\vspace{-5mm}\caption{Mean molecular weight as a function of atmosphere metal mass fraction (\Zatm) for a photosphere with C/O$=0.5$ (blue curve) and a water-depleted photosphere due to cold trapping (orange curve). For miscible-envelope sub-Neptunes, ``warm/hot gas dwarfs'' have mean molecular weights below $<3.3$ largely consistent with that of giant planets. ``Sub-Neptunes with Metal-rich Envelopes'' with $\Zatm>25\%$, the mean molecular weight rapidly increases from the value of 2--3 of H$_2$/He atmospheres to $MMW=13$ at $\Zatm>75\%$, finally approaching that of H$_2$O ($\mu=18$) in the metal-dominated regimes, e.g., for ``steam worlds''. Further increase of the mean molecular weight is possible in the metal-dominated regime as CO$_2$ ($\mu=44$) becomes increasingly abundant for $\Zatm>75\%$. From the bottom up, horizontal gray dashed lines indicate the mean molecular masses of Jupiter (2.2), Neptune (2.53--2.69), and TOI-270\,d ($5.47_{-1.14}^{+1.25}$), with dotted lines indicated the uncertainties.}
\label{fig:Zatm_vs_MMW}
\end{figure*}

Two independent quantities important in setting the interior structure of sub-Neptunes are the the equilibrium temperature ($T_\mathrm{eq}$) and the overall fraction of the planet's mass in volatile metals (C, N, O, S, P), $f_{\mathrm{Zenv}}$. $T_\mathrm{eq}$ is set by the current level of stellar irradiation the planet receives and absorbs, while $f_{\mathrm{Zenv}}$ is set by the planet's formation history. Hence, $T_\mathrm{eq}$ and $f_\mathrm{Zenv}$ are independent parameters. Combined they determine the interior structure and the atmospheric metal mass fraction (\Zatm). We illustrate the temperature dependency for a fixed volatile metal mass fraction ($f_{\mathrm{Zenv}}$) in Figure~\ref{fig:Temp_vs_InteriorStructure}, and we illustrate three scenarios for different ($f_{\mathrm{Zenv}}$) and its effect on the atmospheric metal mass fraction and mean molecular mass for the special case of a sufficiently warm sub-Neptune like TOI-270\,d in Figure~\ref{fig:Zatm_vs_MMW}.

In this Section, we will first discuss the temperature dependency of the interior structures of sub-Neptunes driven by the planet's equilibrium temperature (Section~\ref{sec:class_temp}) and the atmospheric metal mass fraction (Section~\ref{sec:Zatmregimes}). 

\subsection{Temperature-Dependent Interior Structures}\label{sec:class_temp}

We distinguish three different interior structure models, driven by the phase changes of H$_2$O, which is believed to be one of the most abundant high-molecular-weight volatile (HMWV) in most planets. We label these three internal structures as ``Hycean World'' for the coldest sub-Neptunes, ``Stratified mini-Neptune'' for cool mini-Neptunes, and ``Miscible-envelope Sub-Neptunes'' for all warm or hot sub-Neptunes/Neptunes (Figure~\ref{fig:Temp_vs_InteriorStructure}).

\subsubsection{Hycean Worlds (coldest)}

We define the term ``Hycean World'' for the special case of a temperate mini-Neptune that has a thin dry low-mean-molecular-mass, hydrogen-dominated layer ($\Zatm<25\%$) on top of a volatile mantle forming a liquid water ocean at the interface \citep{madhusudhan_habitability_2021,madhusudhan_carbon-bearing_2023}. 
For this to occur, the hydrogen layer must be sufficiently thin for the temperature at the bottom of the hydrogen layer to be sufficiently low for liquid water to form. Likely the planet also has to be very cold, with equilibrium temperatures below approximately 200\,K \citep{pierrehumbert2023runaway}. These planets, if they exist, could potentially be hospitable to life and be highly amenable to characterization via transmission spectroscopy thanks to their low mean molecular weight $\mu<3$ \citep{madhusudhan_habitability_2021}. 

For Hycean worlds, it is however difficult to infer their overall volatile metal mass fraction ($f_{\mathrm{Zenv}}$) because oxygen, as one of the most abundant volatile metals, is condensed below the photosphere, similar to Uranus and Neptune in the Solar System.


\subsubsection{``Stratified Mini-Neptunes'' (cool)}

The term ``mini-Neptune'' has sometimes been used as a synonym for ``sub-Neptune'' in the scientific literature. In this work, however, we define a ``mini-Neptune'' as a specific category of sub-Neptune, which is characterized by a stratified interior structure with a low-metallicity hydrogen-dominated atmosphere on top of a super-critical phase composed of high-molecular-weight volatiles (Figure~\ref{fig:Temp_vs_InteriorStructure}). We argue that this scenario may arise in cool sub-Neptunes such as K2-18b following \citet{innes2023runaway}. H$_2$O and H$_2$ are completely miscible in the supercritical phase at high pressures resulting in a largely homogeneous H$_2$-H$_2$O interior with other high molecular weight volatiles mixed in (e.g., CH$4$ and CO). However, as the condensable H$_2$O is higher molecular mass than H$_2$, inhibited convection may arise when the atmospheric temperature is near the condensation point of H$_2$O. As H$_2$O rains out of the uppermost layers while convection is inhibited, H$_2$O is not transport back up and is efficiently removed from the uppermost layers. This can occur without forming a liquid water ocean on the planet as would be the case for the hycean world scenario (Figure~\ref{fig:Temp_vs_InteriorStructure}). Nonetheless, in the case of the stratified mini-Neptune, the atmospheric metal mass fraction and C/O of the photosphere are also not representative of the bulk envelope, because oxygen is efficiently removed from the photosphere due to H$_2$O condensation.

\subsubsection{``Miscible-envelope Sub-Neptunes'' (warm/hot)}

For most sub-Neptunes known to date, with equilibrium temperatures above approximately 300--350\,K, we argue that temperatures are sufficiently high for H$_2$O not to condense in the envelope and atmosphere. Then, the H$_2$0 and H$_2$ can remain miscible throughout the envelope. H$_2$0 and H$_2$ are in the miscible supercritical phase at high pressures \citep{soubiran_miscibility_2015, pierrehumbert2023runaway,innes2023runaway}, which then directly transitions into the miscible gas phase at lower pressures, where H$_2$ will also remain mixed with H$_2$O in water vapor form. An identifier of sub Neptunes in this regime is a mean molecular mass significantly increased compared to the value of Jupiter and Neptune (2.2 and 2.53–2.69 respectively).

We argue that TOI-270\,d, with an equilibrium of 350--380\,K, is a ``miscible-envelope sub-Neptune''. The high mean molecular weight of $5.47_{-1.14}^{+1.25}$, which is significantly increased compared to the H$_2$-dominated atmospheres of giant planets, is a result of that. The retrieved photospheric temperature of $T_\mathrm{phot}=  385.3_{-41.8}^{+44.2}\,K$, significantly above the water condensation temperature at photospheric pressures, is also consistent with this scenario (Section~\ref{sec:temp_clouds}). In line with that, our GCM modeling also finds that that the temperature remains sufficiently high all around the planet (Section~\ref{sec:cloudfree}, Figure~\ref{fig:gcm_pt}). 

As the vast majority of known sub-Neptunes are likely ``miscible-envelope sub-Neptunes'', we discuss the miscible envelopes for different \Zatm\ regimes in more detail below (Section \ref{sec:Zatmregimes}).

\subsection{\Zatm\ Regimes for Warm/Hot Sub-Neptunes}\label{sec:Zatmregimes}

On warm and hot sub-Neptunes where water cannot condense, we expect a miscible envelope whose atmospheric metal mass fraction (\Zatm) is almost exclusively set by the relative amounts of hydrogen (f$_\mathrm{H}$) and volatile metals (C, N, O, P, S) (f$_\mathrm{CNOPS}$) on the planet. The depth of the envelope is, in turn, set by the total amount of hydrogen and volatile metals relative to the rock/iron mass fraction of the planet f$_\mathrm{rock/iron}$ (see the schematics in Figure~\ref{fig:Zatm_vs_MMW}). In this warm and hot temperature regime of miscible atmospheres, we can broadly distinguish between warm gas dwarfs, sub-Neptunes with metal-rich envelopes, and steam worlds, driven by the planets' appearances in transmission spectroscopy.

\subsubsection{``Warm/Hot Gas Dwarfs''}
We define a ``warm/hot gas dwarf'' as a sub-Neptune-sized planet with a low-metallicity hydrogen-dominated gas envelope and a mean molecular mass of below 3.5 similar to giant planets, or equivalently an atmospheric metal fraction of $\Zatm<25\%$  (Figure~\ref{fig:Zatm_vs_MMW}). For these planets, transmission spectroscopy reveals an atmosphere of maximum pressure scale height due to the low mean molecular weight. Depending on the temperature we expect H$_2$O, CO, CH$_4$ to be the main absorbers in the transmission spectrum of warm/hot gas dwarfs. Despite the relative low metal abundances ($<1\%$), the atmospheric signatures will be large, unless clouds block them.

The warm/hot gas dwarfs scenario is relevant in particular because it can be hard to distinguish from a hycean world or a stratified mini-Neptune in transmission spectroscopy based on the strength of the molecular features alone. This is true in particular for moderate-temperature planets where both scenarios are plausible. The implication for the planet's total amount in volatile metals (f$_\mathrm{CNOPS}$) is, however, drastically different. For a hycean world or a stratified mini-Neptune, the atmospheric metal mass fraction is not representative of the overall envelope because oxygen is condensed out in the form of H$_2$O below the photosphere. For a warm/hot gas dwarf, however, a low \Zatm\ would indicate an overall low volatile-metal content for the planet (f$_\mathrm{CNOPS}$). A robust H$_2$O detection in the photosphere could be a strong indication for warm/hot gas dwarf; however, CH$_4$ absorption can hide the signature of H$_2$O in practice.

\subsubsection{``Sub-Neptunes with Metal-rich Envelopes''}
We define a ``metal-rich sub-Neptune'' as a planet with a largely homogeneous miscible volatile+hydrogen envelope with approximately equal mass fractions in H$_2$/He and high-molecular-weight volatiles (CH$_4$, H$_2$O, CO). Specifically, we define a ``metal-rich sub-Neptune'' to have a \Zatm\ between 25\% and 75\%, resulting in mean molecular weight between approximately 3.5 and 13 (Figure \ref{fig:Zatm_vs_MMW}). The ``metal-rich sub-Neptune'' scenario can be easily identified for sufficiently warm sub-Neptunes via the intermediate value of the mean molecular mass. 

The fact that H$_2$/He and high-molecular-weight volatiles are present in similar abundances mean that the planet's mass must be predominately in rock and iron for sufficiently warm sub-Neptunes with fully miscible envelopes.


\subsubsection{``Steam Worlds''}
Finally, we define ``steam worlds" or ``high-volatile-metal-mass-fraction worlds" as sub-Neptunes with atmospheric metal mass fractions greater than 75\%. These metal-dominated sub-Neptunes can be ``steam worlds" dominated by water vapor, or depending on the atmospheric C/O/N/H ratio, the envelopes of these planets could also be dominated by CH$_4$, H$_2$O/O$_2$, O$_2$, or N$_2$. In that case, we can name them ``Methane worlds", etc. At this level of \Zatm, the main component of the atmosphere critically depends on the overall C/O/N/H ratio, with volatiles like H$_2$O, CH$_4$, CO$_2$, CO, O$_2$, and/or NH$_3$ dominating the mean molecular weight budget ($\mu>13$). Similar to the metal-rich worlds, steam and metal-dominated worlds can have H$_2$ dissolved in the miscible supercritical volatile interior, albeit with H$_2$O, CO, or CH$_4$ dominating the mass budget. In this way, a significant amount of H$_2$/He gas could have been accreted, but it resides in a large homogeneous miscible supercritical interior without creating a hydrogen-dominated or hydrogen-rich atmosphere at the top.
Metal-dominated worlds with $\Zatm>75\%$ can also condense water out of the uppermost layers of their atmospheres if they are sufficiently cold. In that case, large amounts of H$_2$O could result in a liquid water ocean residing between the supercritical interior and a thin atmosphere dominated by all the remaining volatiles such as CH$_4$, N$_2$, NH$_3$, CO$_2$.

\section{Sub-Neptune Chemistry, Climates, and Cloudiness} 

\subsection{Cloud-free Temperate Sub-Neptunes}\label{sec:cloudfree}

TOI-270\,d's largely cloud-free photosphere throughout the near-infrared presents the best SNR spectrum yet to support to the idea that temperate sub-Neptunes ($T_{\mathrm{eq}}<400\,$K) present an enormous opportunity to understand the nature of sub-Neptunes \citep{benneke_water_2019,madhusudhan_carbon-bearing_2023}.
Early studies to determine the compositions of warm sub-Neptunes were plagued by high-altitude clouds or haze that muted the transmission spectroscopy signals \citep[e.g.][]{kreidberg_clouds_2014, knutson_featureless_2014, kempton_reflective_2023}, but increasingly strong evidence is mounting that this effect is mostly restricted to the 400--800\,K temperature range, and that sub-Neptunes with colder equilibrium temperatures are relatively unaffected by high-altitude clouds and hazes \citep{brande_clouds_2024}. This in turn can indicate that temperate sub-Neptunes will have relatively low Bond albedo values, as aerosols can play a major role in reflecting stellar light \citep{charnay_3d_2015,kempton_reflective_2023}. This idea is corroborated by the measured photospheric temperature of TOI-270\,d, which we find is consistent with Bond albedo values in the 0.0--0.3 range, from self-consistent temperature profiles calculated for the retrieved C/H and O/H ratios with SCARLET.
In any case, this work shows that high-precision observations that are sensitive to even medium-MMW and high-MMW atmospheres should be aimed for across the entire sub-Neptune regime, because otherwise transmission spectra may be found to be apparently flat (at low SNR) even in the absence of clouds. 

\begin{figure}[t!]
\begin{center}
\includegraphics[width=1.00\linewidth]{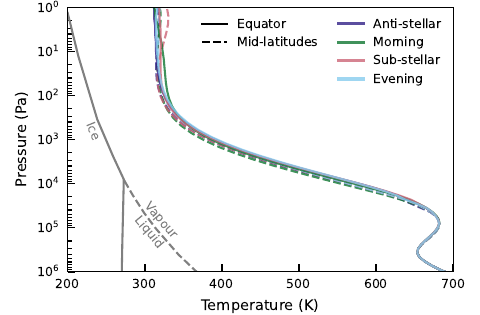}
\end{center}
\vspace{-5mm}\caption{Pressure-temperature profiles from the dual-grey GCM model of TOI-270\,d, averaged over the final 100 days of the simulation. Profiles at the equator (solid lines) and at a latitude of $\theta=45^{\circ}$ are shown. TOI-270\,d is sufficiently warm around the planet that water remains miscible and transitions directly from the vapor form at low pressures into the supercritical phase at high pressure.} 
\label{fig:gcm_pt}
\end{figure}

We also consider the possibility that temperature inhomogeneity may lead to nightside clouds on TOI-270\,d. This was investigated for the cooler K2-18\,b by \citet{charnay_2021} who found water cloud formation for metallicities $\geq 100\times\,\mathrm{solar}$, with a strong dependence on the density of cloud condensation nuclei (CCNs). Due to the inhomogeneous cloud cover, however, the albedo for the planet remained low. For the warmer TOI-270\,d, we use the model outlined in Section \ref{sec:gcm} to investigate the atmospheric structure of the planet. We find that an equatorial jet with a speed of $0.6\,\mathrm{km\,s^{-1}}$ forms and that the transport is sufficient for the atmospheric temperature to remain relatively homogeneous, as shown in Figure \ref{fig:gcm_pt}. The model predicts some variability ($\sim 30\,\mathrm{K}$) at pressures $\lesssim 10^2\,\mathrm{Pa}$; however, the temperatures remain sufficiently high around the planet that clouds are unlikely to form for the water abundances found in the free retrieval.

\subsection{Atmospheric chemistry and the absence of ammonia in TOI-270\,d's photosphere}
\label{sec:atmos_chem}
\begin{figure}[t!]
\begin{center}
\includegraphics[width=1.00\linewidth]{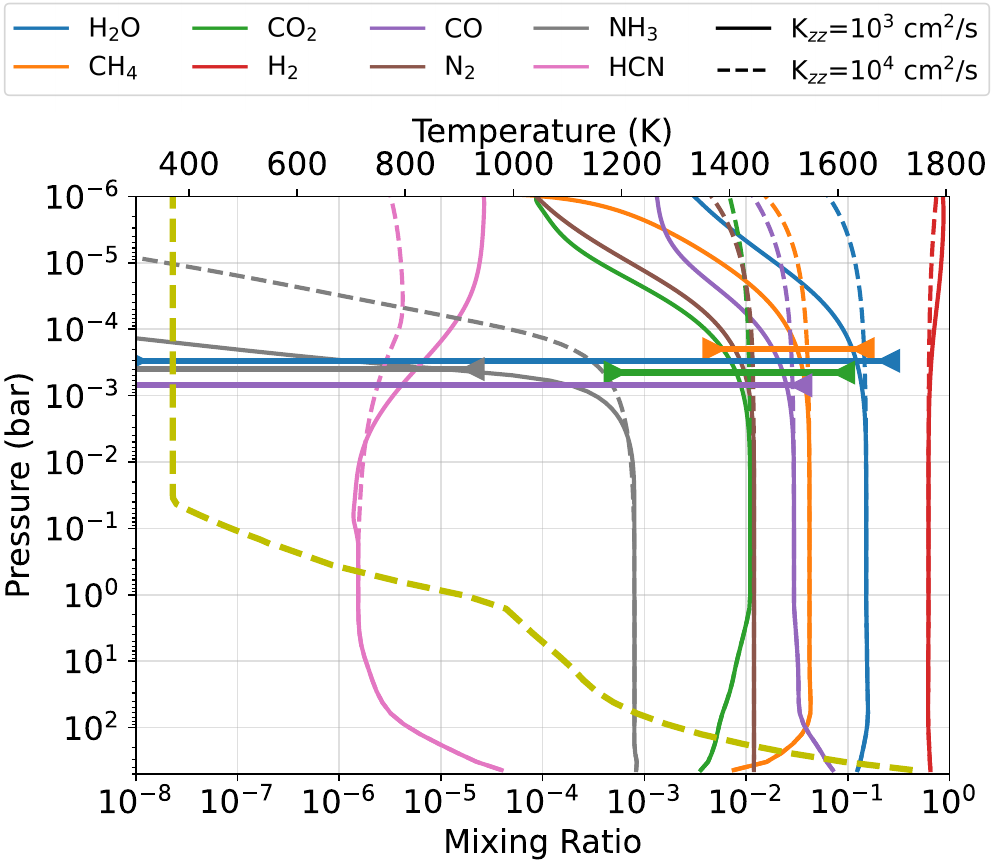}
\end{center}
\vspace{-5mm}\caption{Self-consistent photochemistry in the ``Metal-rich Atmosphere" scenario (metallicity = $230 \times$ solar, solar C/O ratio) reproduces observed molecular abundances. Atmospheric abundances from Photochem model (solid and dashed lines) plotted alongside observed 2-sigma abundances constraints (horizontal solid) lines at approximate pressure levels probed by transits. Model CH$_4$, H$_2$O, CO, and CO$_2$ are consistent with observational constraints independent of vertical mixing, K$_{zz}$. The abundance of NH$_3$ in the photosphere near 0.01-10 mbar is more sensitive to K$_{zz}$ but can be explained by weak vertical mixing, K$_{zz}\leq10^3$ cm$^2$\,s$^{-1}$.} 
\label{fig:Kzz_comparison_updated}
\end{figure}

Using two independent self-consistent atmospheric chemistry models (EPACRIS and Photochem, described in Section~\ref{sec:EPACRIS_method} and Section~\ref{sec:Photochem_method}), we found that a ``metal-rich'' atmosphere ($\sim200\times$ solar, solar C/O ratio) produces the observed molecular abundances. Particularly, thermochemical equilibrium and transport-induced quenching produce the observed abundances of CO$_2$, CH$_4$, and H$_2$O at $\sim1$ mbar. An example of a Photochem run is shown in Figure~\ref{fig:Kzz_comparison_updated}.
Our atmospheric chemistry models predict a higher abundance of CO compared to CO$_2$, while the observations only obtain an upper limit on the CO abundance. However, the relative abundance of CO and CO$_2$ at the quench point is highly sensitive to the local temperature. At $\sim 1$ bar, the temperature in the atmosphere of TOI-270\,d is regulated by the balance between visible light and infrared opacity and can vary based on the assumed Bond albedo (Section~\ref{sec:EPACRIS_method}) and the absorbing species included in the model - which can explain the tension with the retrieval results (Table \ref{tab:detect}). Considering that at the 2$\sigma$ upper limit, the CO abundance could be as high as the few percent predicted by the atmospheric chemistry models, we find that TOI-270\,d is consistent with the ``metal-rich Sub-Neptune'' scenario.

Low abundances of NH$_3$ have previously been proposed as possible evidence against a ``metal-rich atmosphere'' \citep{yu2021identify,hu_unveiling_2021,tsai2021inferring,madhusudhan_chemical_2023}, but it has also been recognized that the abundance of NH$_3$ at the pressure probed by transmission spectra is also controlled by a competition between vertical mixing and photodissociation \citep{hu2021photochemistry}. While our retrieval analysis shows that ammonia is largely absent from the photosphere ($<$ 50 ppm), there are numerous plausible explanations for this non-detection. Weak vertical mixing (tropospheric K$_{zz}<$10$^3$ cm$^2$\,s$^{-1}$) can produce low abundances of the photochemically fragile NH$_3$ while not substantially modifying CO$_2$, CH$_4$, and H$_2$O abundances (Figure \ref{fig:Kzz_comparison_updated}).
Additionally, atmospheric nitrogen inventories that are under-abundant relative to C and O could further contribute to low NH$_3$ abundances in the detectable part of the atmosphere. A low atmospheric N inventory is a plausible outcome of reducing magma ocean conditions, which preferentially causes N to partition into dissolved phases \citep{shorttle_distinguishing_2024}. Because TOI-270\,d is unlikely to be a temperate Hycean world given its high equilibrium temperature and the fact that both self-consistent 1D modeling and 3D GCMs predict temperatures too high for the cold-trapping of water (Figure \ref{fig:gcm_pt}), the non-detection of NH$_3$ strengthens the argument that the absence of detectable NH$_3$ is possible in ``metal-rich atmospheres'' and does not necessarily require a liquid-water ocean.


With a sulfur chemistry network, EPACRIS also predicts \ce{SO2} with a mixing ratio of $\sim2$ ppm at 1 mbar, increasing to $\sim20$ ppm at 10 mbar, supporting the retrieved abundance of SO$_2$ in the upper atmosphere (Table~\ref{tab:detect}). This SO$_2$ is produced by atmospheric photochemistry, with the photolysis of H$_2$O providing the oxidizing agent. Meanwhile, our atmospheric chemistry model can only produce CS$_2$ with a mixing ratio of $10^{-10}$ near 1 mbar, and instead suggest substantial formation of OCS in the upper atmosphere. This behavior is at odds with the free retrieval results.
It is possible that unidentified or missing photochemical pathways lead to the production of \ce{CS2}. An example could be the photoexcitation of carbon monoxide in the upper atmosphere, potentially leading to enhanced \ce{CS2} production through the formation of reduced carbon species from the reaction CO* + CO $\rightarrow$ C + \ce{CO2} \citep{Yang-2023}, which might subsequently react with sulfur-bearing molecules.

\begin{figure}[t!]
\begin{center}
\includegraphics[width=1.0\linewidth]{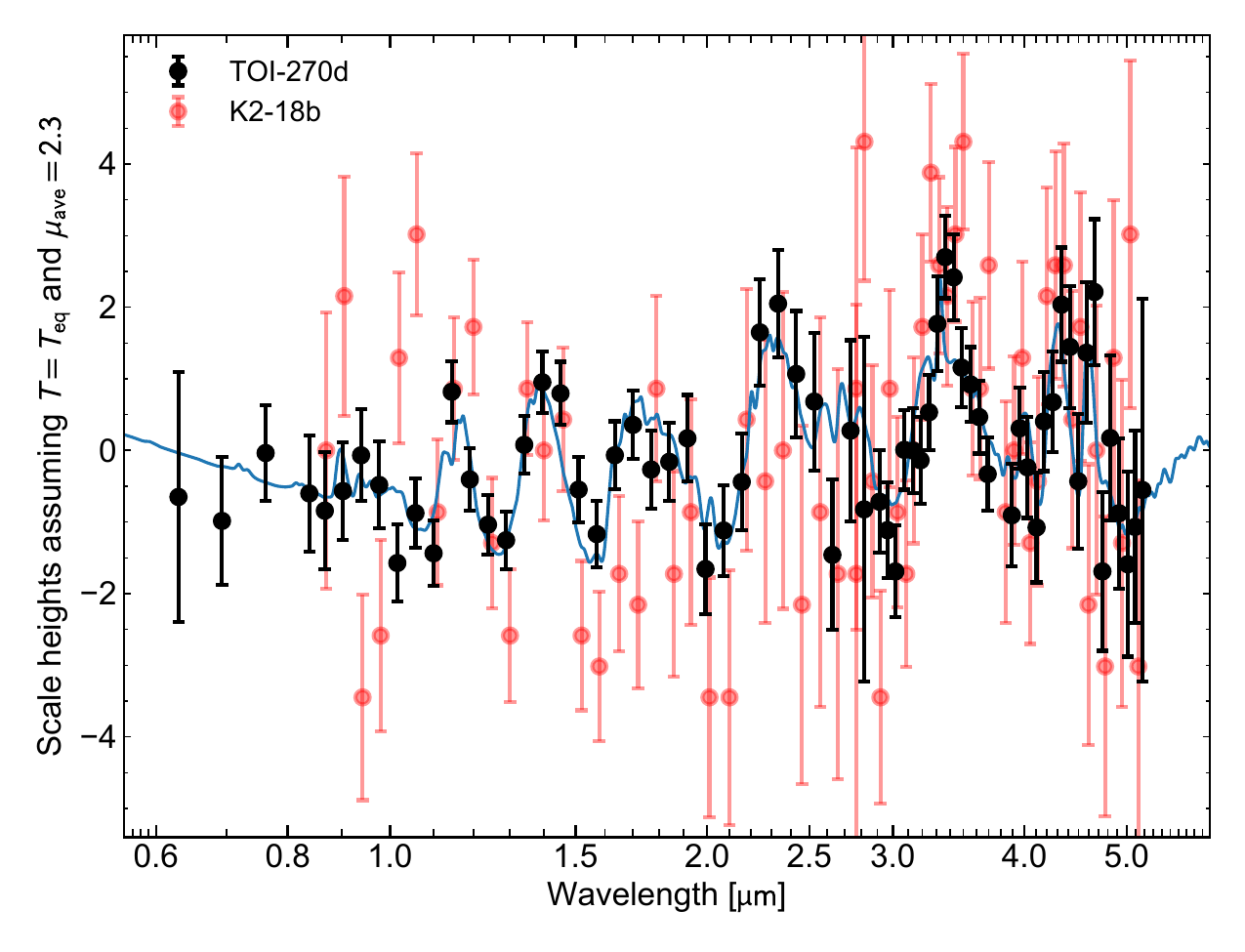}
\end{center}
\vspace{-5mm}\caption{Empirical comparison of transmission spectra of TOI-270\,d (black point, this work) and K2-18\,b \citep[red points][]{madhusudhan_carbon-bearing_2023}. The blue curve represents the best-fitting model from the free-chemistry retrieval for TOI-270\,d (see also Figure \ref{fig:TransmissionSpectrum}). The TOI-270\,d observations provide 2.5 smaller error bars in terms of atmospheric scale height, thanks to the higher Transmission Spectroscopy Metrics of TOI-270\,d (TSM=97) vs K2-18\,b (TSM=42). 6 SOSS and 6 G395H transits of K2-18\,b would be needed to reach the same sensitivity to atmospheric characterization for K2-18b, and the smaller atmospheric features due to the large mean molecular weight of TOI-270\,d's would likely not have been detectable for K2-18b.} 
\label{fig:TOI270d_vs_K2_18b}
\end{figure}

\subsection{TOI-270\,d, K2-18\,b, and the Hycean scenario}



While the similarities between TOI-270\,d and K2-18\,b are numerous, as both planets are sub-Neptunes with low equilibrium temperatures, as well as detections of CH$_4$ and CO$_2$ in their upper atmospheres, the small differences that exist between the two planets make them quite different, both from an observational and compositional perspective. Observationally, TOI-270\,d's TSM is 2.5 times higher than K2-18\,b's, resulting in error bars 2--3 times smaller relative to the atmospheric scale height for TOI-270\,d, regardless of the atmosphere scenario (Figure~\ref{fig:TOI270d_vs_K2_18b}). This explains how our single-transit observations with NIRISS and NIRSpec were able to precisely constrain the atmosphere composition of TOI-270\,d, down to a precise scale height measurement that provided us with the MMW and temperature of the photosphere.

In addition, the difference in planet radius and mass of TOI-270\,d and K2-18\,b also results in different constraints on the types of envelopes they can host. At 2.6$\,\re$ and 8.6$\,\me$ \citep{benneke_water_2019}, K2-18\,b is almost two times more massive than TOI-270\,d and can more easily retain a low-MMW, thin H$_2$/He envelope against atmospheric escape. This is why K2-18\,b is not usually discussed as a ``water world" or ``steam world" candidate, as it is the case for the smaller and two-times-less-massive TOI-270\,d \citep[e.g, discussion in][]{roy_water_2023}. 

Finally, and potentially most importantly, despite only a small difference in equilibrium temperature ($<100\,$K), TOI-270\,d and K2-18\,b appear to fall into different temperature regimes with regard to the phase transitions of water (Section~\ref{sec:class_temp}). Similarly to K2-18\,b, TOI-270\,d has been proposed as a candidate ``hycean world'' \citep{madhusudhan_habitability_2021, rigby_ocean_2024}; however, planetary climate models with self-consistent treatments of water vapor and cloud feedback indicate that such a state would require the atmosphere to be very thin ($<0.1-1$ bar) and the planet to have a high Bond albedo ($>0.7-0.9$) \citep{innes2023runaway,leconte20243d}. In Section~\ref{sec:EPACRIS_method}, we simulated the expected pressure-temperature profiles of this scenario using EPACRIS and found that the surface temperature would be $\sim400$ K for a 0.1-bar H$_2$-dominated atmosphere with a Bond albedo of 0.7 and $\sim430$ K for a 1-bar H$_2$-dominated atmosphere with a Bond albedo of 0.9. 

In summary, the slightly higher equilibrium temperature of TOI-270\,d makes this planet theoretically less likely to be Hycean in nature than K2-18\,b \citep{innes2023runaway, pierrehumbert2023runaway, leconte20243d}, which is in line with our observations. The high Bond albedo values needed to produce liquid water conditions on TOI-270\,d are inconsistent with the observed photosphere temperature which suggests a low Bond albedo around 0.0--0.3. Instead, in its slightly warmer regime, TOI-270\,d likely hosts water in a supercritical state, and thus points at a miscible and well-mixed atmosphere that can explain the high abundances of CH$_4$, CO$_2$ and H$_2$O, as well as the higher photosphere temperature simultaneously. Thus, K2-18\,b and TOI-270\,d provide us with an opportunity to study the transition in sub-Neptune compositions around the water condensation temperature, with K2-18\,b being in the hycean or stratified mini-Neptune and TOI-270\,d being in the miscible-envelope sub-Neptune regime (Figure~\ref{fig:Temp_vs_InteriorStructure}). Further observations of both targets, as well as new atmosphere analyses of similar sub-Neptunes, will be vital in understanding sub-Neptune compositions and the possibility for liquid water across this temperate regime.



\section{Summary and Conclusions}\label{conclusions}

We observed the full 0.6--5.2$\mum$ transmission spectrum of the $2.2\re$ exoplanet TOI-270\,d using \textit{JWST NIRISS/SOSS} and \textit{JWST NIRSpec/G395H} delivering a sensitivity to atmospheres previously unseen in the sub-Neptune regime (Figure~\ref{fig:TransmissionSpectrum}). In particular, the detection of five molecular absorption features of CH$_4$ (9.4$\sigma$) across the near-infrared enables us to uniquely constrain the mean molecular weight of TOI-270\,d's atmosphere \citep{benneke_atmospheric_2012}. Intriguingly, we find a mean molecular weight of $5.47_{-1.14}^{+1.25}$ for the envelope of TOI-270\,d, significantly elevated when compared to the hydrogen-dominated atmospheres of Jupiter and Neptune with mean molecular weights of 2.2 and 2.53--2.69, respectively (Figure \ref{fig:retrieval}).

Consistent with the high mean molecular weight, we derive the atmospheric metal mass fraction to be $\Zatm=58_{-12}^{+8}\%$, meaning that approximately half the mass in the atmosphere is in volatile metals. Carbon-, oxygen-, and sulfur-bearing species can directly be seen in the transmission spectrum, with other volatile elements such as nitrogen (in the form of $N_2$) impacting only the mean molecular weight.

The high atmospheric metal mass fraction of TOI-270\,d demonstrates that sub-Neptune envelopes need not have a distinct low-metallicity H$_2$/He layer on top of a denser H$_2$O mantle or rock/iron core. Instead, we classify TOI-270\,d as a “miscible-envelope sub-Neptune” hosting a metal-rich envelope with H$_2$/He and the high-molecular-weight volatiles mixed into a largely miscible water vapor + supercritical envelope (Section~\ref{sec:classification}). This agrees with recent theoretical work that predicted miscible envelope for sufficiently warm sub-Neptunes \citep{pierrehumbert2023runaway,innes2023runaway}. We find that TOI-270\,d is in a sweet-spot in terms of equilibrium temperature, for which the envelope is warm enough for water vapor not to rain out of the photosphere, and cool enough for the atmosphere to be virtually cloud-free and amenable to atmospheric characterization. This gives a hitherto unique view into the atmospheric metal mass fraction and C/O of a planet in the sub-Neptune regime.


\subsection{TOI-270\,d as a rocky planet that accreted H$_2$/He}

Our measurement of the atmosphere metal mass fraction provides us with the critical third measurable to break the degeneracy in the mass-radius diagram. Measurement of the planetary radius and mass generally deliver degenerate results for a sub-Neptune's bulk composition because low-density water-dominated interiors with a small hydrogen envelope can inherently lead to the overall same bulk density as a denser rock-dominated interior to a larger hydrogen envelope \citep[e.g.,][]{rogers_three_2009}. For fully-miscible metal-rich envelopes, the atmospheric metal mass fraction (\Zatm) becomes the critical third measurable that, together with the measured planet mass and radius, enables the constraint of the total amount of mass in rock/iron, H$_2$/He, and volatile metals (C, N, O, etc.). 

Taking into account a fully miscible interior, we obtain a rock/iron mass fraction of $90_{-4}^{+3}$\,wt\,\% rock/iron for TOI-270\,d (Section~\ref{sec:Rock_Iron_Core_Mass}), suggesting that TOI-270\,d is almost completely a rocky planet by mass. We show that TOI-270\,d is a rocky planet that formed interior to the ice line (dry formation) and accreted a few wt \% of H$_2$/He. The high abundance of CH$_4$, CO$_2$, and H$_2$O could then be the result of high-temperature magma-ocean/envelope reactions that delivered an atmosphere with similar mass fraction of H$_2$/He and high-molecular-weight volatiles, without the need for any accretion of ices from the protoplanetary disk. TOI-270\,d could be an archetype of a close-in sub-Neptune, that may all similarly have rock-dominated interiors. This is in disagreement with the standard ``ice giant'' picture of Neptune and Uranus's interior structure, but instead confirms the implications derived from atmospheric escape modeling studies that seem to best match the observed planet demographics and position of the radius valley under the assumption of predominantly rocky interiors.\citep[e.g., ][]{fulton_california-kepler_2018, owen_atmospheric_2019, gupta_sculpting_2019}.

\subsection{Earth and Venus in light of TOI-270\,d and sub-Neptunes}
The ramifications of work related on sub-Neptunes is not limited to distant exoplanets. Instead, sub-Neptunes may also provide critical insights into the origin of water and other volatiles on the terrestrial planets in the Solar System \citep{young_earth_2023}. Our work indicates that TOI-270\,d is likely a $4.8\,\me$ rock-dominated planet, potentially a scaled up version of a terrestrial planet, that accreted a few wt\,\% of H$_2$/He. The estimated $0.33_{-0.13}^{+0.22}\,\me$ of high-molecular-weight volatiles (H$_2$O, CH$_4$, CO, CO$_2$) in TOI-270\,d's envelope were then plausibly not accreted externally, but as we show, may have potentially formed from high-temperature magma-ocean/hydrogen reactions \citep{schlichting_chemical_2022}. The resulting amount of high-molecular-weight volatiles corresponds to $1244_{-464}^{+761}$ times the mass of Earth’s oceans on a $4.28_{-0.47}^{+0.46}\,\me$ rocky/iron planet. \citet{young_earth_2023} suggested that magma-ocean reactions may have similarly formed the H$_2$O on Earth, leaving behind Earth's oceans when the hydrogen envelope was lost to space. Unlike the Earth, however, TOI-270\,d was able to hold onto its envelope likely due to its deeper gravitational potential, leaving it in its original state with both the H$_2$/He and the volatiles produced from magma-ocean envelope reaction still in its envelope and available for detailed characterization. TOI-270\,d may therefore present an opportunity to better understand the geochemical processes at the magma-ocean interface that may also have led to the formation of Earth's oceans as well as the water on other rocky planets including Early-Venus. It also raises the intriguing possibility that many rocky planets with oceans or water envelopes much deeper than the Earth can exist in an intermediate mass regime \citep[e.g., ][]{piaulet_evidence_2023}, resulting from magma-ocean/envelope reactions with a temporarily accreted H$_2$/He envelope that is subsequently lost to space. 

\subsection{Uranus and Neptune in light of TOI-270\,d}
Our image of TOI-270\,d as a rocky planet that accreted H$_2$/He is in contrast to our standard image of Neptune and Uranus, which are generally labeled as the ``ice giants'' in the Solar System, implying a volatile-dominated interior resulting from accretion of ice-rich material. The ``ice giant'' label makes sense from a formation perspective, given that Uranus and Neptune are situated well beyond the Solar System's water ice line, making available large amounts of icy material for the rapid accretion of ice-dominated planet cores. The fact that they host a H$_2$-dominated envelope indicates that Uranus and Neptune formed quickly, within 10~Myr, before the dissipation of the protoplanetary gas disk. Nonetheless, it is worth noting that even for Uranus and Neptune, the debate regarding their interior composition is not fully resolved \citep[][and reference therein]{teanby_neptune_2020}. Current observations of Uranus and Neptune can be fit similarly well with a wide range of interior models including ice-dominated and rock-dominated interiors (ice giants vs rock giants), and an interior with more rock than ice is more compatible with the ice sources available in the outer solar system \citep{teanby_neptune_2020}. Making conclusive statements for Uranus and Neptune remains a challenge, in particular, because overall O/H ratio cannot be inferred reliably \citep[e.g.,][]{luszcz-cook_constraining_2013}. Future studies of sub-Neptune exoplanets in the context of Uranus and Neptune may help develop a more complete picture of the formation of planets in the sub-Neptune-to-Neptune regime, especially as conclusive measurement of the interior rock-to-ice ratio of Uranus and Neptune will be hard to obtain in the near future, while many warm sub-Neptune exoplanet will be observed in increasing detail in the years to come.


\vspace{10mm}
We would like to acknowledge all the attendees of the ``Density Matters 2024'' Ringberg Workshop, in particular the organizers Rafael Luque and Remo Burn as well as Peter Gao, Nikku Madhusudhan, Natalie Batalha, James Owen, William Misener, Thomas Henning, and many more for the valuable discussions on subjects related to this work. This work is based on observations with the NASA/ESA/CSA James Webb Space Telescope, obtained at the Space Telescope Science Institute (STScI) operated by AURA, Inc. All of the data presented in this paper were obtained from the Mikulski Archive for Space Telescopes (MAST) at the Space Telescope Science Institute. The specific observations analyzed can be accessed via \dataset[10.17909/dvqh-2r48]{https://doi.org/10.17909/dvqh-2r48}.  B.B. and P.-A.R. acknowledge financial support from the Natural Sciences and Engineering Research Council (NSERC) of Canada. P.-A.R. further acknowledges support from the University of Montreal, and from the Trottier institute for exoplanets (iREx). C.P. acknowledges support from the NSERC Vanier scholarship, and iREx. B.B. also acknowledges financial support from the Canadian Space Agency and the Fond de Recherche Québécois-Nature et Technologie (FRQNT; Québec). This work has been carried out within the framework of the NCCR PlanetS supported by the Swiss National Science Foundation under grants 51NF40\_205606.

\clearpage

\bibliography{refs_Benneke, refs_coauthors_add_here}{}
\bibliographystyle{aasjournal}

\end{document}